\documentclass[a4paper,fleqn,usenatbib]{mnras}

\usepackage{txfonts}

\usepackage[T1]{fontenc}
\usepackage{ae,aecompl}
\usepackage{graphics}

\usepackage{psfig}   
\usepackage{graphicx}
\usepackage{amssymb}
\usepackage{subfigure}

\title[]{On the origin of planetary-mass objects in NGC\,1333}

\author[R.~J.~Parker \& C.~Alves~de~Oliveira]{
  Richard J.~Parker\thanks{E-mail: R.Parker@sheffield.ac.uk}\thanks{Royal Society Dorothy Hodgkin Fellow}$^1$ and Catarina Alves de Oliveira$^2$ \vspace*{0.1cm}\\
   $^1$Department of Physics and Astronomy, The University of Sheffield, Hicks Building, Hounsfield Road, Sheffield, S3 7RH, UK \\
   $^2$European Space Agency (ESA), European Space Astronomy Centre (ESAC), Camino Bajo del Castillo s/n, 28692 Villanueva de la Ca{\~n}ada, \\ Madrid, Spain}

\begin{document}

                             
\pagerange{\pageref{firstpage}--\pageref{lastpage}} \pubyear{2023}

\maketitle

\label{firstpage}

\begin{abstract}
 The dominant formation mechanism of brown dwarfs and planetary mass objects in star-forming regions is presently uncertain. Do they form like stars, via the collapse and fragmentation of cores in Giant Molecular clouds, or do they form like planets in the discs around stars and are ejected via dynamical interactions?  In this paper, we quantify the spatial distribution of substellar objects in NGC\,1333, in particular focusing on planetary-mass objects that have been the target of recent deep imaging observations. We find that these objects have a spatial distribution that is indistinguishable from the stars, and more massive brown dwarfs. We also analyse $N$-body simulations and find that a population of ejected planets would have a significantly different spatial and kinematic distribution to stars, and brown dwarfs that also formed through gravitational collapse and fragmentation.  We therefore conclude that the low-mass substellar objects in NGC\,1333 formed more like stars than planets, although we predict that a population of hitherto undetected ejected planetary mass objects may be lurking in this, and other star-forming regions.
\end{abstract}


\section{Introduction}

Star and planet formation occur contemporaneously, yet they are often treated as distinct or separate processes. This simplification becomes unviable when assessing the substellar population in star-forming regions. Observations show that there is one brown dwarf ($m<0.08$\,M$_\odot$) for every $\sim$2--7 H-burning stars  in star-forming regions \citep[e.g.][]{Barrado02,Andersen08,Geers11,Muzic15,Pearson20,Kubiak21}, and most authors consider the brown dwarf regime an extension of the same process that formed stars with a continuous mass function into the substellar regime \citep[e.g.][]{Chabrier14}, though see \citet{Thies08}.

The origin of free-floating planetary mass objects is even less clear, especially as their masses are notoriously difficult to determine \citep[e.g.][]{Baraffe02,Feiden12,Canty13,Lueber22}, and the mass range of these objects overlaps with brown dwarfs \citep{Esplin17,Gagne17,Lodieu21}. Therefore, a population of free-floating planets in a star-forming region may have simply formed ``like stars'', i.e.\,\,from the gravitational collapse and fragmentation of Giant Molecular Clouds \citep[e.g.][]{Low76,Padoan04,Gahm07,Haworth15}; or could instead be the result of planet--planet scattering \citep[e.g.][]{Chatterjee08,Boley12,Veras12,Smullen16}, or direct encounters between two stars, leading to the ejection of planets around one or both stars \citep{Bonnell01b,Parker12a,DaffernPowell22}.

Recently, \citet{Scholz22} used the numbers of free-floating planets from simulations to predict how many such objects could be observed with James Webb Space Telescope. Following this, \citet{Scholz23} performed deep imaging on substellar objects in NGC\,1333 to determine whether these objects hosted discs. \citet{Scholz23} established that only one out of the six least massive PMOs in NGC\,1333 hosts a disc, leading them to speculate that these objects may have formed more like planets, rather than like stars.

Given the prospect of detailed high-resolution spectra with JWST \citep[e.g\,\,the NIRSpec instrument,][]{Jakobsen22}, which will enable accurate mass and velocity determinations of substellar objects, it is prescient to determine what -- if any -- signatures in the spatial and kinematic distribution of substellar objects could be used to pinpoint their physical origins.

Previous work has shown that if brown dwarfs are a continuous extension of star formation, we would not expect a significantly different spatial distribution of the brown dwarf population \citep{Parker14c}. However, a similar analyses of the distribution of ejected planets has not been performed.

In this paper, we exploit the likely complete census of stars and brown dwarfs in the NGC\,1333 star-forming region \citep{Luhman16} to calculate the spatial distribution of brown dwarfs and planetary-mass objects, and how this compares to the stars. We then calculate the same metrics in $N$-body simulations whose initial conditions were derived from a previous analysis of the spatial distribution of stars in NGC\,1333 \citep{Parker17a}. However, these simulations differ from those in \citet{Parker17a} in that -- in addition to a continuous IMF into the brown dwarf mass regime -- they contain substellar objects on orbits around stars, which may be ejected from their host stellar system due to dynamical encounters.

The paper is organised as follows. In Section~2 we describe the observational data we used to calculate the spatial distributions of stars, brown dwarfs and planetary mass objects in NGC\,1333. In Section~3 we describe our methods to measure the spatial distributions, and also to set-up the $N$-body simulations. In Section~4 we present our results, in Section~5 we provide a discussion, and we conclude in Section~6. 

\section{Observational data}

We use the same dataset as in \citet{Parker17a}, which in turn is based on the \citet{Luhman16} census of NGC\,1333. \citet{Luhman16} confirmed spectroscopically the membership of tens of brown dwarfs and stars, resulting in a final sample of 203 stars.

\citet{Scholz23} performed deep imaging on a subset of 14 of the brown dwarfs from \citet{Luhman16}, as well as an additional substellar object discovered by \citet{Esplin17}. We add this object to our sample, and use the spectral types to estimate the masses of stars earlier than M0 using the temperature scale from \citet{Schmidt82}, the scale from \citet{Luhman03b} for sources between M0 and M9.5 and for L dwarfs we used the scale from \citet{Lodieu08}.

The masses were derived in the same way as in \citet{Parker17a}, assuming a distance of 235\,pc to NGC\,1333.  Using this method, the masses of the L0 objects were set to 0.012\,M$_\odot$, the L1 objects were set to 0.010\,M$_\odot$ and the L3 object was assigned a mass of 0.05\,M$_\odot$ in our subsequent analysis. Note that our results are not dependent on the exact mass values, but rather the relative masses, and we assume that the L-type objects are the lowest mass members of NGC\,1333.

\begin{figure}
\begin{center}
\rotatebox{270}{\includegraphics[scale=0.35]{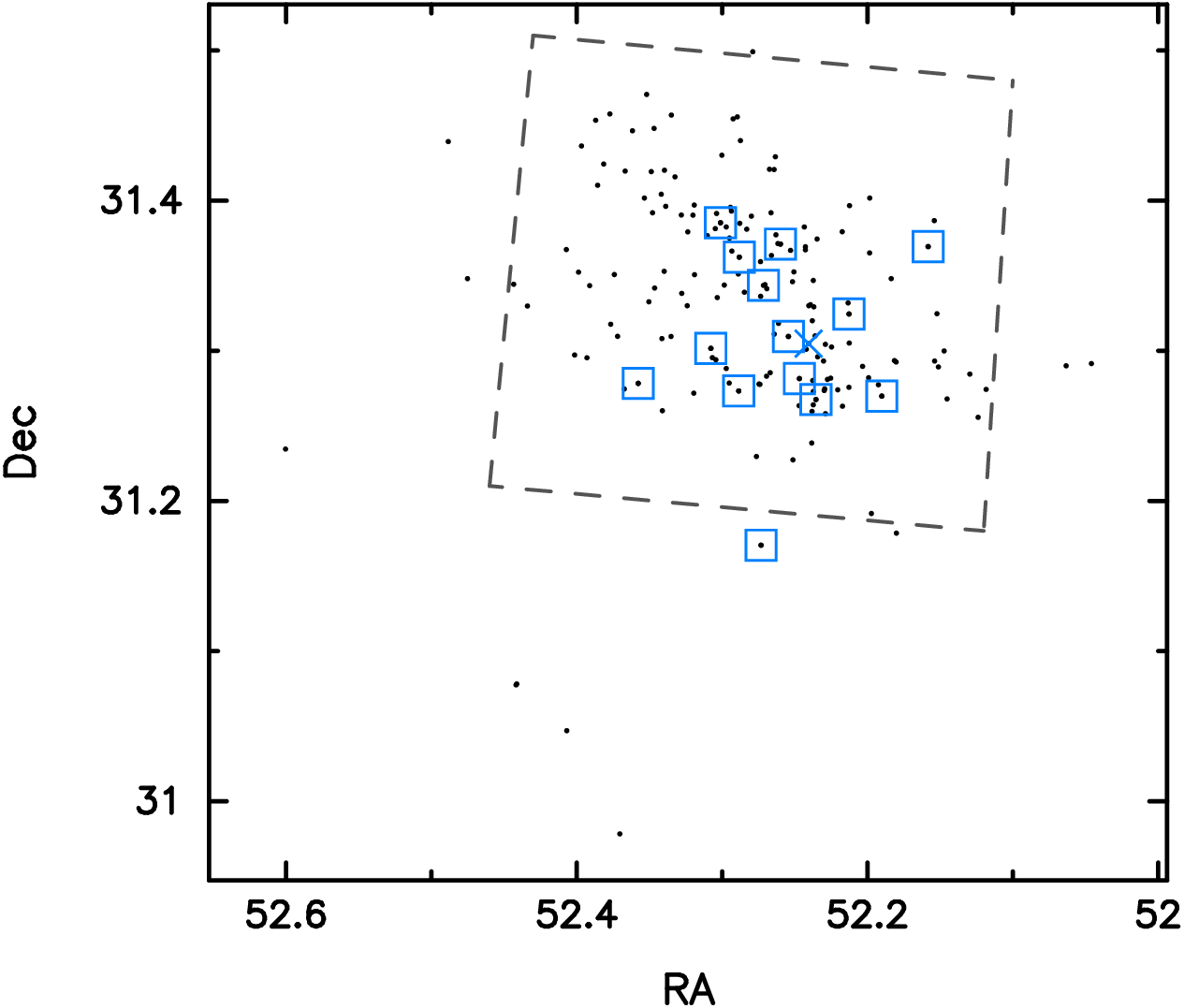}}
\end{center}
\caption[bf]{Map of objects in NGC1333. The area within the dashed lines is the ACIS-I field from \citet{Luhman16}, which is observationally complete. The 15 planetary-mass objects discussed in \citet{Scholz23} are shown in blue, with the object not in the original \citet{Luhman16} census shown by the blue cross \citep{Esplin17}.}
\label{ngc1333_map}
\end{figure}

We show the positions of the objects in the \citet{Luhman16} sample in Fig.~\ref{ngc1333_map}. The 15 planetary mass objects discussed in \citet{Scholz23} are shown by the blue symbols, with the new object from \citet{Esplin17} shown by the blue cross. The census of NGC\,1333 is thought to be complete within the ACIS-I field, as discussed in \citet{Luhman16}, and this is shown within the dashed line. 

\section{Methods}
\label{method}

In this section we first describe our methods to quantify the spatial and kinematic distributions of stars, brown dwarfs and planetary mass objects, before describing the $N$-body simulations with which we compare the observations. 

\subsection{Quantifying the spatial distributions of objects}

There are multiple methods in the literature for quantifying the spatial distribution of stars and substellar objects in star-forming regions \citep[see][for a discussion of the different methods]{Parker15b,BlaylockSquibbs22}. We utilise two different techniques to quantify the spatial distributions of stars, brown dwarfs and planetary mass objects in the observed census of NGC\,1333, and our $N$-body simulations, $\Lambda_{\rm MSR}$ \citep{Allison09a} and $\Sigma - m$ \citep{Maschberger11}.

An enormous amount of confusion abounds in the literature when assessing the advantages and disadvantages of different techniques for quantifying spatial distributions, including the amount of mass segregation, in a star-forming region. Often, misunderstandings and apparent contradictions occur because mass segregation is often defined in different ways.

For example, $\Lambda_{\rm MSR}$, which we will describe below, measures whether a subset of objects are closer together than a randomly chosen subset. $\Sigma - m$ measures the relative local surface densities of the objects in a chosen subset. Typically, a smooth, centrally concentrated and mass-segregated star cluster will show mass segregation in both $\Lambda_{\rm MSR}$ and high surface densities of the most massive objects in $\Sigma-m$. However, in an association, with multiple groups, or nodes, of stars, the massive stars may be as spread out as a randomly chosen subset, yet they may be in areas of higher than average surface density \citep[and such behaviour often occurs in associations due to dynamical evolution,][]{Parker14b}.

$\Lambda_{\rm MSR}$ and $\Sigma-m$ measure different properties, but both have the advantage that the subset of interest can be any group of objects, as defined by the user. Often, we are interested in the subset of the most massive stars, but in this paper we are interested in planetary mass objects and brown dwarfs. The strength of $\Lambda_{\rm MSR}$ and $\Sigma-m$ is that they are relative measures, and are not hamstrung by e.g. the absence of massive stars, as erroneously asserted by \citet{Guszejnov22}.

\subsubsection{The $\Lambda_{\rm MSR}$ mass segregation ratio}

The mass segregation ratio, $\Lambda_{\rm MSR}$ is calculated by constructing a minimum spanning tree (MST), a graph of the shortest possible path between a set of points, or nodes, where there are no closed loops \citep{Prim57}. We construct an MST between the chosen subset and calculate the length of this MST, $l_{\rm subset}$, which contains $N_{\rm MST}$ objects. We then calculate the average MST length in the star-forming region by taking a set of 100 randomly chosen MSTs, each containing the same  number of objects as the chosen subset, and calculating the average MST length from this, $\langle l_{\rm average} \rangle$. We conservatively estimate the lower (upper) uncertainty as being the length that lies 1/6 (5/6) through an ordered list of the random subset lengths, corresponding to a 66\,per cent deviation from the random length, $\langle l_{\rm average} \rangle$. This is summarised in the following equation:
\begin{equation}
  \Lambda_{\rm MSR} = {\frac{\langle l_{\rm average} \rangle}{l_{\rm subset}}} ^{+ {\sigma_{\rm 5/6}}/{l_{\rm subset}}}_{- {\sigma_{\rm 1/6}}/{l_{\rm subset}}}.
  \label{lambda_msr}
\end{equation}
If a subset of objects is mass-segregated (i.e. closer together than the average subset), then $\Lambda_{\rm MSR} >> 1$. If the objects in the chosen subset are more spread our than the average objects in the region (as might be expected for planetary-mass objects or brown dwarfs) then  $\Lambda_{\rm MSR} << 1$. If no mass segregation is present, $\Lambda_{\rm MSR} = 1$.

There are two ways of determining $\Lambda_{\rm MSR}$ to assess the significance of any deviation from unity. The original method in \citet{Allison09a} starts with the $N_{\rm MST}$ most massive objects, and then calculates  $\Lambda_{\rm MSR}$ for successively larger $N_{\rm MST}$ values. \citet{Allison09a} used this method to show that  the four most massive stars in the Orion Nebula Cluster (i.e.\,\,the Trapezium system) are mass-segregated, but the amount of mass segregation decreases with larger $N_{\rm MST}$ such that when the $N_{\rm MST} = 20$ most massive objects are considered, there is no mass segregation (by definition $\Lambda_{\rm MSR} = 1$ when $N_{\rm MST}$ includes all the stars in the region). This version of $\Lambda_{\rm MSR}$ has been used to quantify mass segregation (or inverse mass segregation) of massive stars, brown dwarfs and pre-stellar clumps/cores  \citep{Moeckel09a,Moeckel09b,Olczak11,Parker11b,Girichidis12,Plunkett18,Hetem19,Konyves20,Morii23}. In this paper, we will apply this method to the five \emph{least} massive objects in NGC\,1333, and then add successively higher-mass objects to $N_{\rm MST}$.

An alternative method, first proposed by \citet{Parker11b} and since used by other groups \citep[e.g.][]{Alfaro16,Gonzalez17,Alfaro18}, keeps $N_{\rm MST}$ fixed and instead slides through the dataset. For example, one can start with the 10 least massive objects, calculate $\Lambda_{\rm MSR}$, and then move to the 11 - 20 least massive objects, and so on. This method is noisier than the original method, and care must be taken to avoid over-interpreting significant deviations from $\Lambda_{\rm MSR} = 1$, but it has the advantage that a specific subset of objects in the middle of the mass range can be examined in detail. This will be important later when we analyse $N$-body simulations with a population of planetary-mass objects that lie in the middle of a wider mass distribution of substellar objects.

\subsubsection{$\Sigma - m$ relative surface densities}

The $\Sigma - m$ technique \citep{Maschberger11} quantifies the relative surface density of objects in a star-forming region, and can then be used to determine whether a particular mass range have higher or lower surface densities compared to the region as a whole. For example, mass segregation of the most massive stars might be apparent in higher-than-average surface densities for these objects. Conversely, if substellar objects are preferentially ejected  over low-mass stars, then we might expect substellar objects to have lower-than average surface densities.

The surface density of an object of mass $m$ is calculated using 
\begin{equation}
\Sigma = \frac{N - 1} {\pi r_{N}^2},
\end{equation}
where $r_N$ is the distance to the $N^{\rm th}$ nearest neighbour to the object of mass $m$ \citep{Casertano85}. We adopt $N = 10$, but the dependence of $\Sigma$ on the choice of $N$ only becomes important if the structure of the region changes abruptly between different values of $N$. In practice, as long as $N$ is higher than e.g. 2 or 3, then the determination of $\Sigma$ is not biased by multiplicity \citep{Kraus08} and traces the typical local surface density of the stellar and substellar systems.

\citet{Maschberger11} showed that in simulated star-forming regions where the most massive stars form in the most dense regions, the massive stars have a significantly (as determined by a two-sided KS test) higher surface densities than the region as a whole. The method was further utilised by \citet{Kupper11} and \citet{Parker14b}, who showed that high surface densities in the massive stars can be used as a dynamical clock  (in tandem with other metrics including $\Lambda_{\rm MSR}$) to determine the initial conditions of a star-forming region. 

\subsection{Velocity distributions}

We do not have information on the velocities of the stars in our observational sample, but we can make predictions for the expected velocity distributions from our $N$-body simulations. We construct two distributions. First, we take the radial velocities, defined in the simulations as the component of the velocity vector along the $z$-axis. Secondly, we produce a distribution of proper motion velocities $\mu$, where we take the on-sky positions (i.e. in the $xy$-plane in the simulations) between snapshots (at $t_n$ and $t_{n+1}$) and then divide the difference in position by the time interval, thus:
\begin{equation}
\mu = \frac{r_{xy, t_{n+1}} - r_{xy, t_n}}{t_{n+1} - t_n}.
  \end{equation}

\subsection{$N$-body simulations}

In order to compare the observed spatial distributions of substellar objects in NGC\,1333 to models of brown dwarf and planet formation, we use $N$-body simulations to simulate the dynamical evolution of this star-forming region. We create populations of $N = 150$ stellar and substellar objects by drawing masses from a \citet{Maschberger13} Initial Mass Function (IMF) with a probability distribution of the form
\begin{equation}
p(m) \propto \left(\frac{m}{\mu}\right)^{-\alpha}\left(1 + \left(\frac{m}{\mu}\right)^{1 - \alpha}\right)^{-\beta}.
\label{maschberger_imf}
\end{equation}
In Eqn.~\ref{maschberger_imf}  $\mu = 0.2$\,M$_\odot$ is the scale parameter, or `peak' of the IMF \citep{Bastian10,Maschberger13}, $\alpha = 2.3$ is the \citet{Salpeter55} power-law exponent for higher mass stars, and $\beta = 1.4$ describes the slope of the IMF for low-mass objects. We randomly sample this distribution in the mass range 0.001 -- 50\,M$_\odot$, such that we sample objects down to the planetary mass regime.

We then create a separate population of planetary-mass objects, which we place on an orbit around stellar mass ($0.08 < m/{\rm M_\odot} \leq 3$) objects. These `planets' have a mass of 10\,M$_{\rm Jup}$ ($9.4 \times 10^{-3}$\,M$_\odot$, which overlaps with the mass range of the objects that form ``like stars''), are assigned zero eccentricity and inclination. In one set of simulations the planets are all assigned a semimajor axis of 5\,au (to be on a Jupiter-like orbit), and in another set of simulations the planets are all assigned a semimajor axis of 30\,au (to be on a Neptune-like orbit). In a third set of simulations, the planets are again all placed at 30\,au, but have masses of 1\,M$_{\rm Jup}$ ($9.4 \times 10^{-4}$\,M$_\odot$, which is slightly lower than the mass range of the objects that form as stars).

We thus have a collection of systems, which are either single stars, single brown dwarfs or star--planet systems. We randomly distribute these systems within a fractal distribution \citep{Goodwin04a}, which is designed to mimic the filamentary and substructured stellar distributions in both observed \citep{Cartwright04,Sanchez09,Hacar13,Buckner19} and simulated \citep{Schmeja06,Bate09} star-forming regions. We refer the interested reader to \citet{DaffernPowell20} for a comprehensive description of the set-up of the fractal distributions, but briefly summarise them here.

The fractals are constructed by placing a parent particle at the centre of a cube, and then determining the probability of that particle maturing and spawning further particles. The probability of this occurring goes as $2^{3-D}$, where $D$ is the desired fractal dimension. For a smooth distribution, the fractal dimension is $D = 3.0$ and so no further particles are spawned. For a substructured distribution, $D = 1.6$, which results in multiple generations of particles. The particles are assigned a velocity drawn from a Gaussian distribution of mean zero. The child particles inherent their parents' velocities, plus a small random offset that decreases with each subsequent generation of particles.

We scale the velocities of the systems to a subvirial ratio ($\alpha = 0.3$, where $\alpha = T/|\Omega|$ and $T$ and $|\Omega|$ are the total kinetic and potential energies, respectively, and $\alpha = 0.5$ is virial equilibrium).  

In our simulations, we mainly adopt a highly substructured distribution ($D = 1.6$) with a radius $r_F = 0.5$\,pc, which results in high stellar densities ($\sim10^4$\,M$_\odot$\,pc$^{-3}$). These initial conditions are informed by earlier work to constrain the initial conditions of NGC\,1333 \citep{Parker17a}, albeit towards the high end of the initial conditions as constrained by \citet{Parker17a}. However, as we want to test whether the PMOs in NGC\,1333 might be the result of ejection from systems, we adopt these high densities as they more readily lead to the creation of a separate population of free-floating planetary mass objects, whose properties (spatial and velocity distribution) we can compare with that of the brown dwarfs in the simulations. However, we also ran sets of simulations with larger (1\,pc) radii and higher fractal dimensions ($D = 2.0$, which are less substructured) resulting in densities of $\sim 500$\,M$_\odot$\,pc$^{-3}$ and found no differences to our main results, save for fewer stars being ejected overall. 

The simulations are evolved for 10\,Myr using the \texttt{kira} package within the Starlab environment \citep{Zwart99,Zwart01} and we analyse the data at 1 and 3\,Myr, which spans the likely mean age range for this star-forming region. We do not include stellar evolution in the simulations.  We summarise the different simulations in Table~\ref{simulations}. In our analysis of the spatial and velocity distributions in the $N$-body simulations, we exclude any objects that are beyond a radius of 5\,pc, in order to mimic an observer's field of view. 

  \begin{table}
  \caption[bf]{A summary of the different initial conditions of our simulated star-forming regions. The columns show the fractal dimension, $D$, the initial radius of the star-forming region, $r_F$, the initial median local stellar density as a result of $D$ and $r_F$, $\tilde{\rho}$, the mass of the planets, $m_p$, and the initial semimajor axis of the planets $a_p$. The final column indicates whether the simulation is shown in a Figure in Section~\ref{results}. }
  \begin{center}
    \begin{tabular}{|c|c|c|c|c|c|}
      \hline

      $D$ & $r_F$ & $\tilde{\rho}$ & $m_p$ & $a_p$ & Fig. \\
      \hline
      1.6 & 0.5\,pc & $10^4$\,M$_\odot$\,pc$^{-3}$ & 10\,M$_{\rm Jup}$ & 30\,au & Figs.~\ref{sim_Lambda}, \ref{sim_sigma-m}~and~\ref{sim_velocities} \\
      1.6 & 0.5\,pc & $10^4$\,M$_\odot$\,pc$^{-3}$ & 1\,M$_{\rm Jup}$ & 30\,au & Fig.~\ref{sim_JN_Lambda} \\
      \hline
      1.6 & 0.5\,pc & $10^4$\,M$_\odot$\,pc$^{-3}$ & 10\,M$_{\rm Jup}$ & 5\,au & --- \\
      \hline

      2.0 & 0.5\,pc & 500\,M$_\odot$\,pc$^{-3}$ & 10\,M$_{\rm Jup}$ & 30\,au & ---\\
      
      2.0 & 0.5\,pc & 500\,M$_\odot$\,pc$^{-3}$ & 1\,M$_{\rm Jup}$ & 30\,au & ---\\

      \hline
    \end{tabular}
  \end{center}
  \label{simulations}
\end{table}

\section{Results}

We first show the results of the $\Lambda_{\rm MSR}$ and $\Sigma - m$ analyses of the observational data for NGC\,1333 before describing the similar analyses performed on $N$-body simulations with substellar objects that form in the same way as stars, and also substellar objects that were initially orbiting a star.

\subsection{NGC\,1333}

\begin{figure*}
  \begin{center}
\setlength{\subfigcapskip}{10pt}
\hspace*{-1.5cm} \subfigure[$\Lambda_{\rm MSR}$, low masses]{\label{NGC1333_Lambda-a}\rotatebox{270}{\includegraphics[scale=0.3]{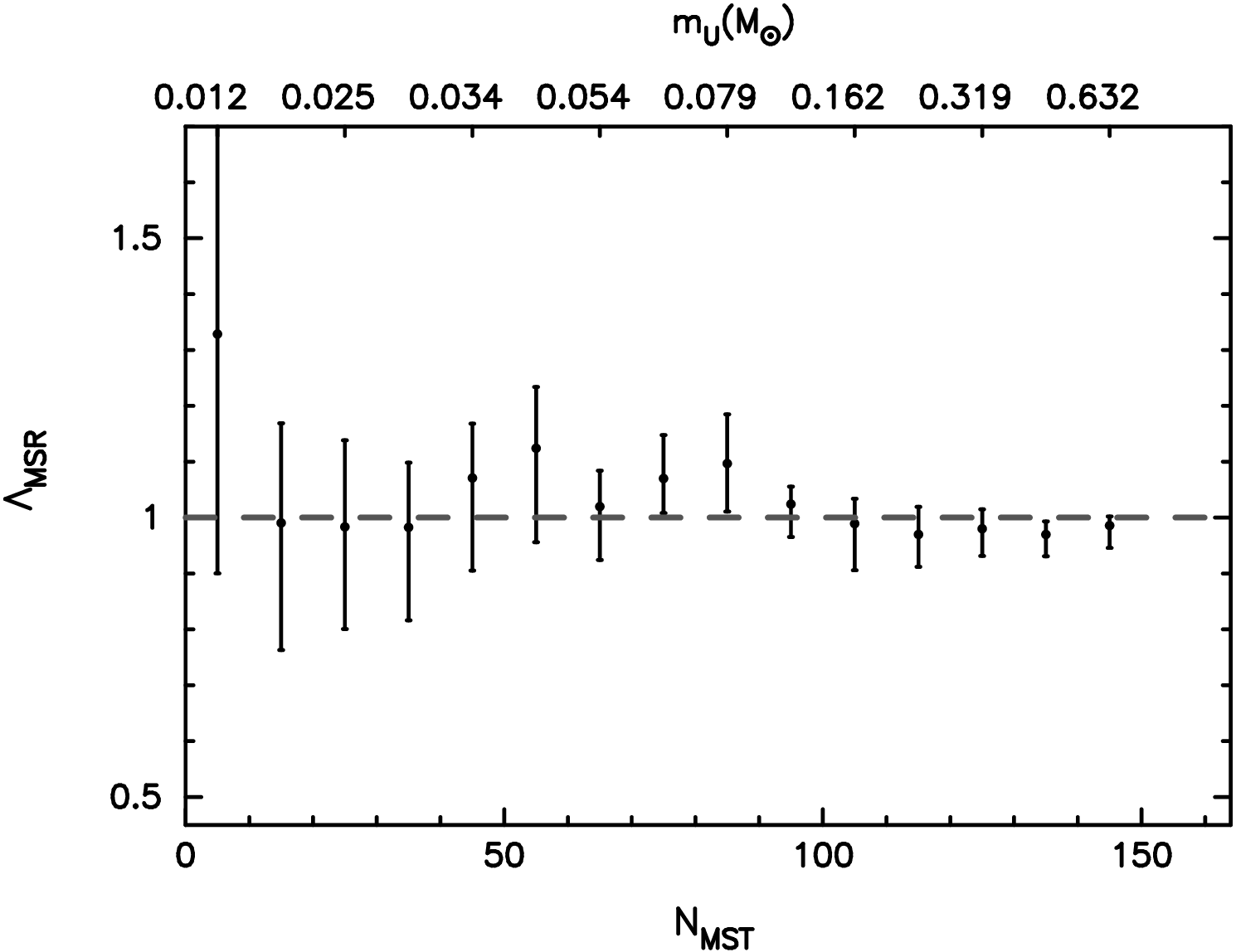}}} 
\hspace*{0.3cm}\subfigure[$\Lambda_{\rm MSR}$, slide]{\label{NGC1333_Lambda-b}\rotatebox{270}{\includegraphics[scale=0.32]{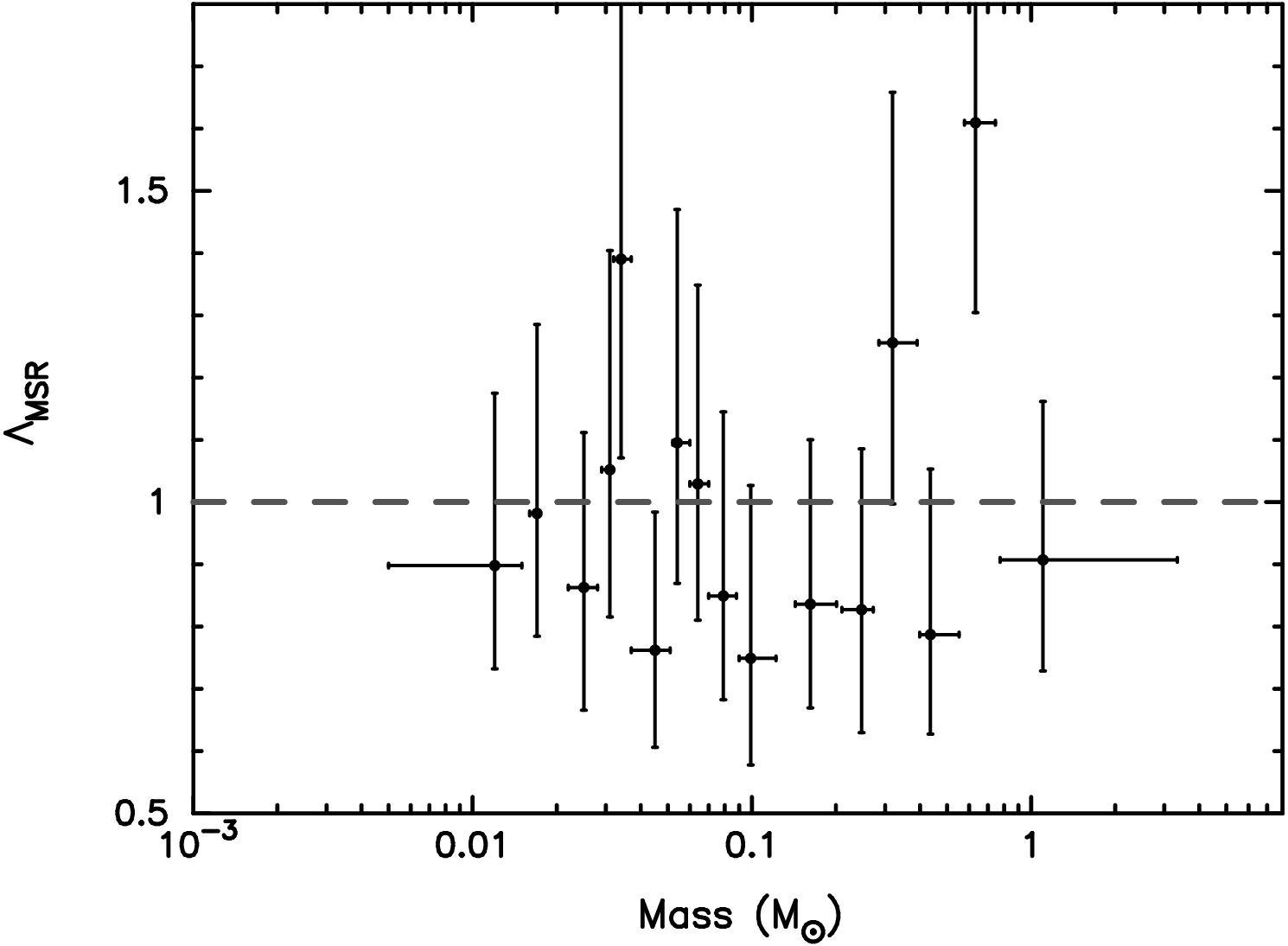}}}
\caption[bf]{Calculation of the spatial distribution of brown dwarfs and planetary mass objects in NGC\,1333 with $\Lambda_{\rm MSR}$. Panel (a) shows the results when calculating $\Lambda_{\rm MSR}$ with the $N_{\rm MST} = 5$ least massive objects, and then subsequently adding the next 10 least massive objects. The mass of the highest mass object within the $N_{\rm MST}$ subset is indicated on the top horizontal axis. Panel (b) shows the calculation of $\Lambda_{\rm MSR}$ for subsets of 10 objects, and moving through the data. In both panels, the error bars in the vertical direction indicate the 1/6 and 5/6 percentile values in the distribution of the randomly chosen MSTs. In panel (b) the `error' bars in the horizontal direction show the mass range for each calculation of $\Lambda_{\rm MSR} $. $\Lambda_{\rm MSR} = 1$, which indicates no mass segregation, is shown by the dashed grey line in both panels.}
\label{NGC1333_Lambda}
  \end{center}
\end{figure*}

We show the calculation of $\Lambda_{\rm MSR}$ for NGC\,1333 in Fig.~\ref{NGC1333_Lambda}. In panel (a) we show the calculation of $\Lambda_{\rm MSR}$ for successively larger values of $N_{\rm MST}$, where progressively higher mass stars are added to the determination of  $\Lambda_{\rm MSR}$. The values for $\Lambda_{\rm MSR}$ are all consistent with unity, i.e.\,\,there is no preferential spatial distribution for the substellar objects and in particular we highlight that there is no difference in the spatial distribution of the least massive PMOs discussed in \citet{Scholz23}.

We then employ the `slide' version of $\Lambda_{\rm MSR}$ in panel (b). As discussed in Section~3, this method is noisier, but allows us to identify different spatial distributions of objects in a narrow mass range (the horizontal `error' bars in panel (b) show the mass range that the value of $\Lambda_{\rm MSR}$ is associated with). Again, we see no significant deviation from unity in any mass range.

\begin{figure}
\begin{center}
\rotatebox{270}{\includegraphics[scale=0.35]{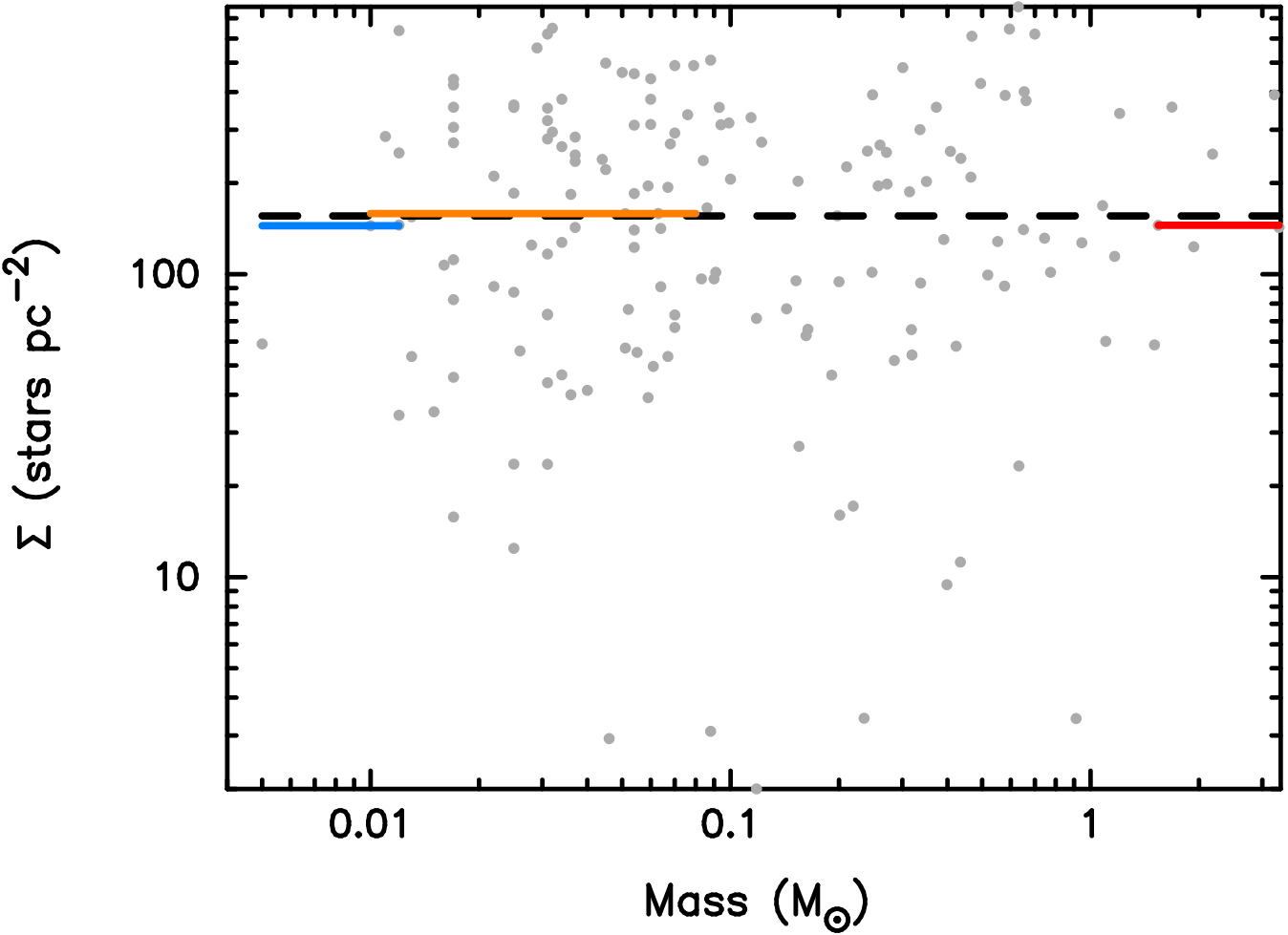}}
\end{center}
\caption[bf]{$\Sigma - m$ plot for NGC\,1333. The surface density of each object, $\Sigma$ is plotted against its mass $m$. The median surface density for the star-forming region is shown by the dashed black line, the median surface density for the most massive stars is shown by the solid red line. The median surface density for brown dwarfs ($0.01 < m/{\rm M_\odot} < 0.08$) is shown by the solid orange line, and the median surface density for the planetary mass objects is shown by the solid blue line. No mass regime/object type has significantly different densities from the region as a whole.    }
\label{ngc1333_sigma-m}
\end{figure}

We next plot the local surface density around each object $\Sigma$ as a function of the object's mass $m$ in Fig.~\ref{ngc1333_sigma-m}. The various horizontal lines represent the median surface densities of subsets of objects; the black dashed line is the median surface density for all objects, the solid red line is the most massive stars, the orange line is for brown dwarfs ($0.01 < m/{\rm M_\odot} < 0.08$) and the solid blue line is for the planetary mass objects discussed in \citet{Scholz23}. Two-sided KS tests between the different subsets and the region as a whole produce high p-values, meaning we cannot reject the hypothesis that all types of object share the same underlying density distribution.\\

The absence of any difference in the spatial distribution of substellar objects is very similar to previous results obtained for NGC\,1333 with a very similar dataset \citep{Parker17a}, but provides an interesting null-result with which to compare our $N$-body simulations.

\subsection{$N$-body simulations}
\label{results}

We run 10 versions of the each simulation, where we alter the random number seed that sets the masses, positions and velocities. However, in the following we show the results for a representative simulation, and if necessary discuss any differences between the different runs.

\begin{figure}
\begin{center}
\rotatebox{270}{\includegraphics[scale=0.35]{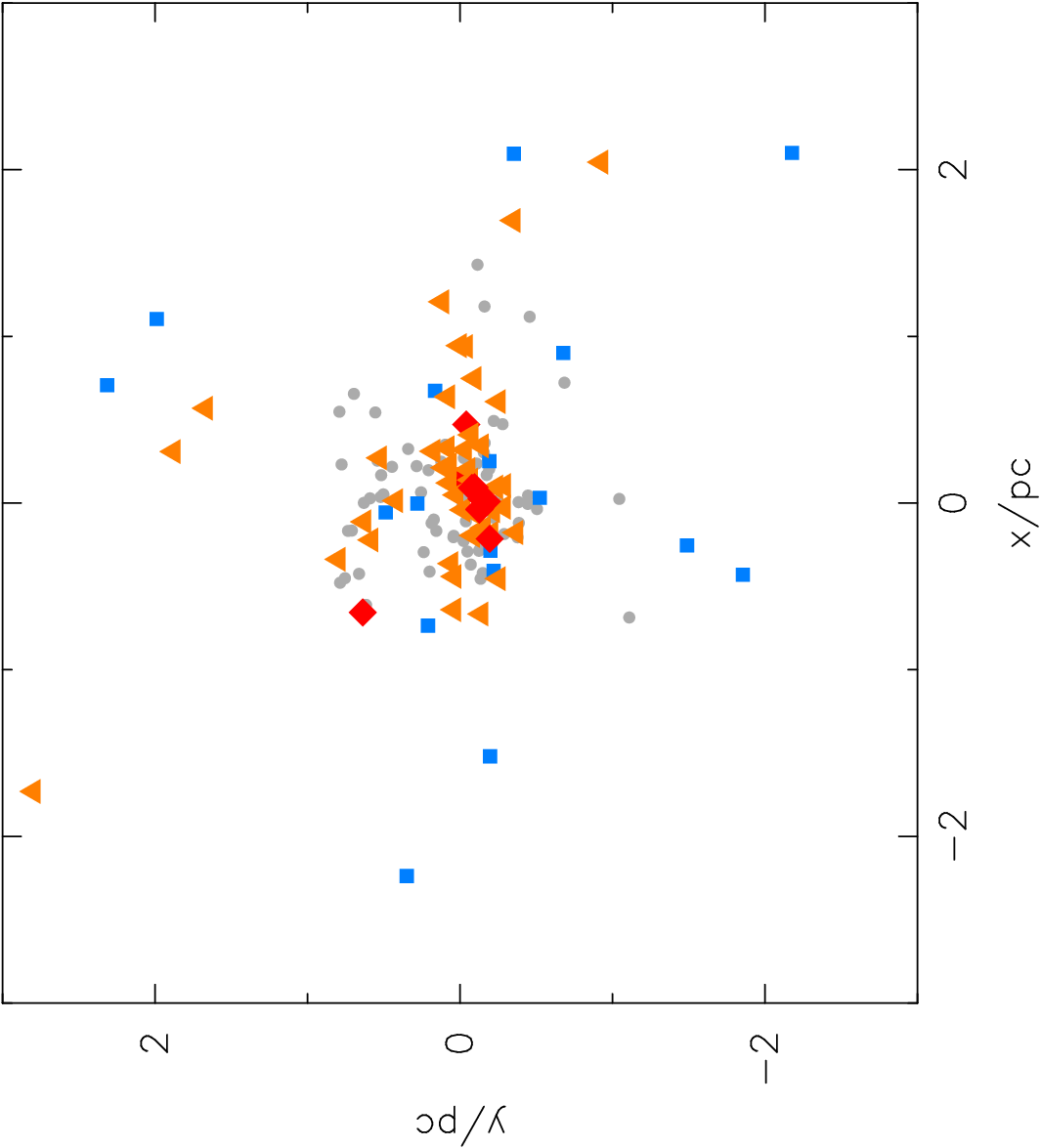}}
\end{center}
\caption[bf]{A plot of the positions of objects at 3\,Myr in a representative simulation. Stars are shown by the grey points (the most massive are shown by the red diamonds), and brown dwarfs are shown by the orange triangles. The ejected planets are shown by the blue squares. }
\label{sim_map}
\end{figure}

In Fig.~\ref{sim_map} we show the positions of stars (grey points), brown dwarfs (orange triangles) and ejected planets (blue squares) after 3\,Myr of dynamical evolution (similar to the age of NGC\,1333). We also show the locations of the most massive stars by the red diamond symbols. From inspection, the planets and brown dwarfs appear more dispersed than the stars, but we quantify this in the following analysis.

\begin{figure*}
  \begin{center}
    \setlength{\subfigcapskip}{10pt}
\hspace*{-1.5cm} \subfigure[$\Lambda_{\rm MSR}$, low masses]{\label{sim_Lambda-a}\rotatebox{270}{\includegraphics[scale=0.32]{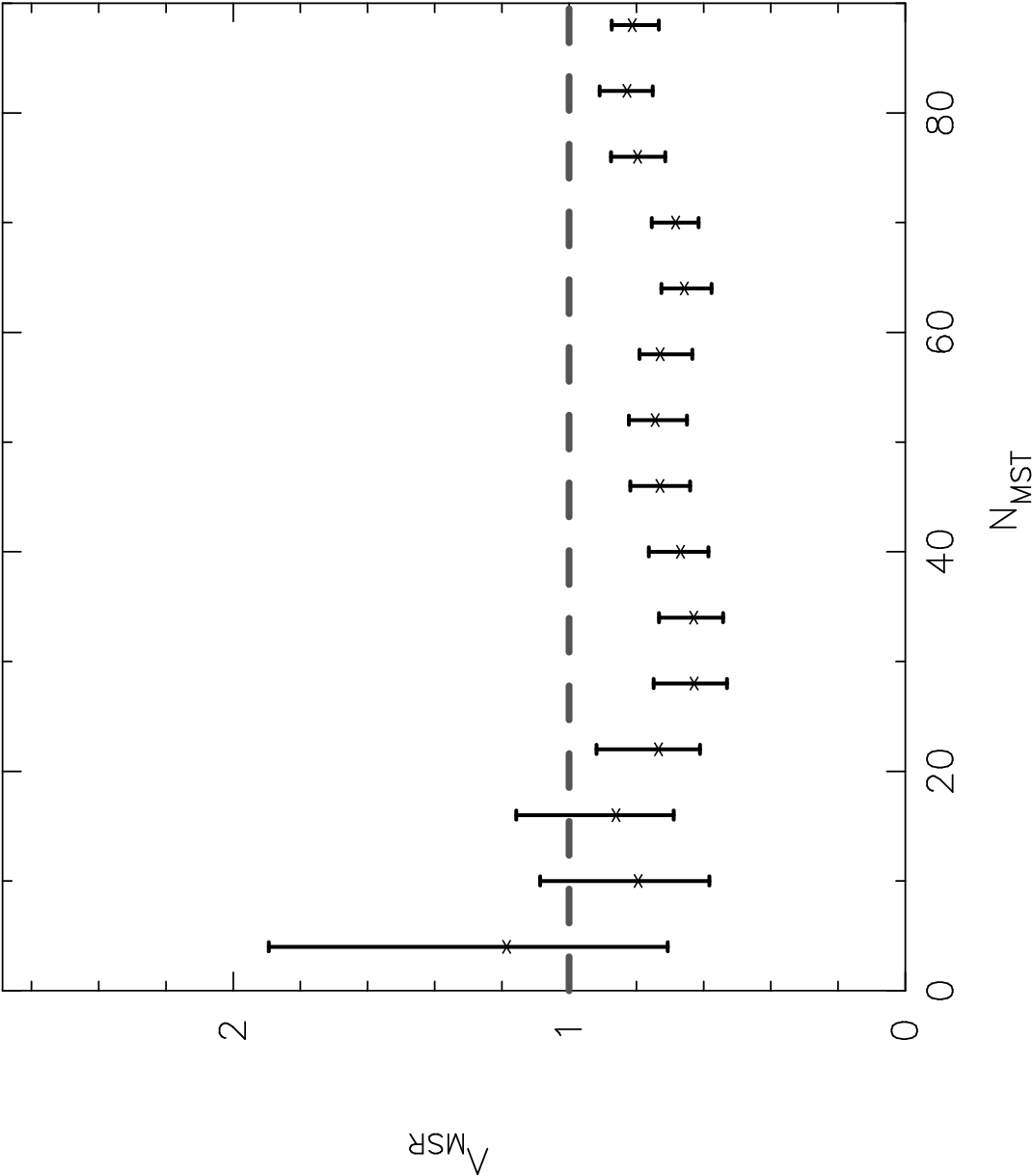}}}
\hspace*{0.3cm}\subfigure[$\Lambda_{\rm MSR}$, slide]{\label{sim_Lambda-b}\rotatebox{270}{\includegraphics[scale=0.32]{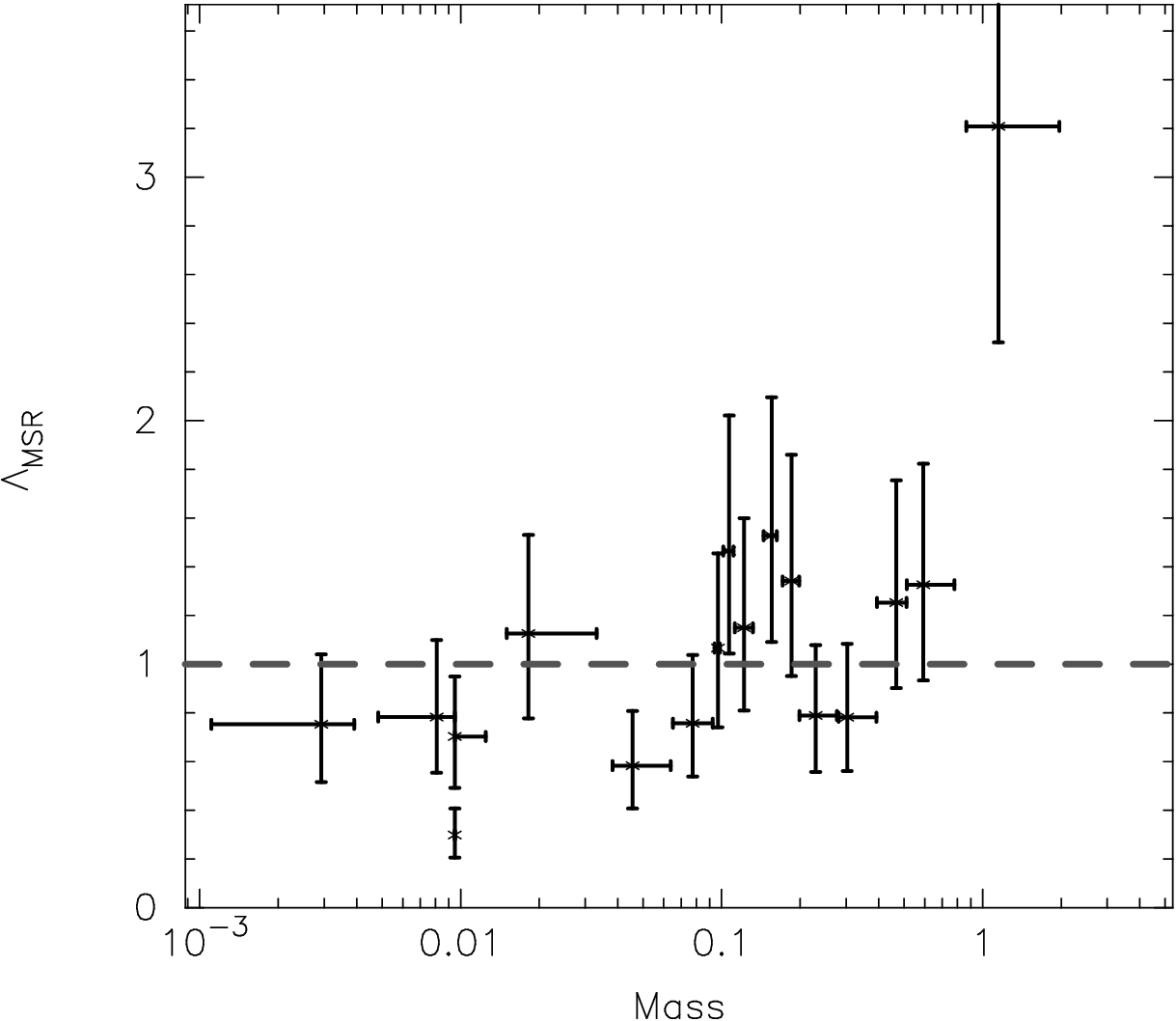}}}
\caption[bf]{Calculation of the spatial distribution of brown dwarfs and planetary mass objects in our simulations with $\Lambda_{\rm MSR}$. Panel (a) shows the results when calculating $\Lambda_{\rm MSR}$ with the $N_{\rm MST} = 5$ least massive objects, and then subsequently adding the next 10 least massive objects. 
  Panel (b) shows the calculation of $\Lambda_{\rm MSR}$ for subsets of 10 objects, and moving through the data. In both panels, the error bars in the vertical direction indicate the 1/6 and 5/6 percentile values in the distribution of the randomly chosen MSTs. In panel (b) the `error' bars in the horizontal direction show the mass range for each calculation of $\Lambda_{\rm MSR} $. $\Lambda_{\rm MSR} = 1$, which indicates no mass segregation, is shown by the dashed grey line in both panels. The bin containing the ejected planetary-mass objects has a mass segregation ratio $\Lambda_{\rm MSR} = 0.35$, and is centred on a mass value of $9.4 \times 10^{-3}$M$_\odot$.}
\label{sim_Lambda}
  \end{center}
\end{figure*}

\begin{figure}
\begin{center}
\rotatebox{270}{\includegraphics[scale=0.35]{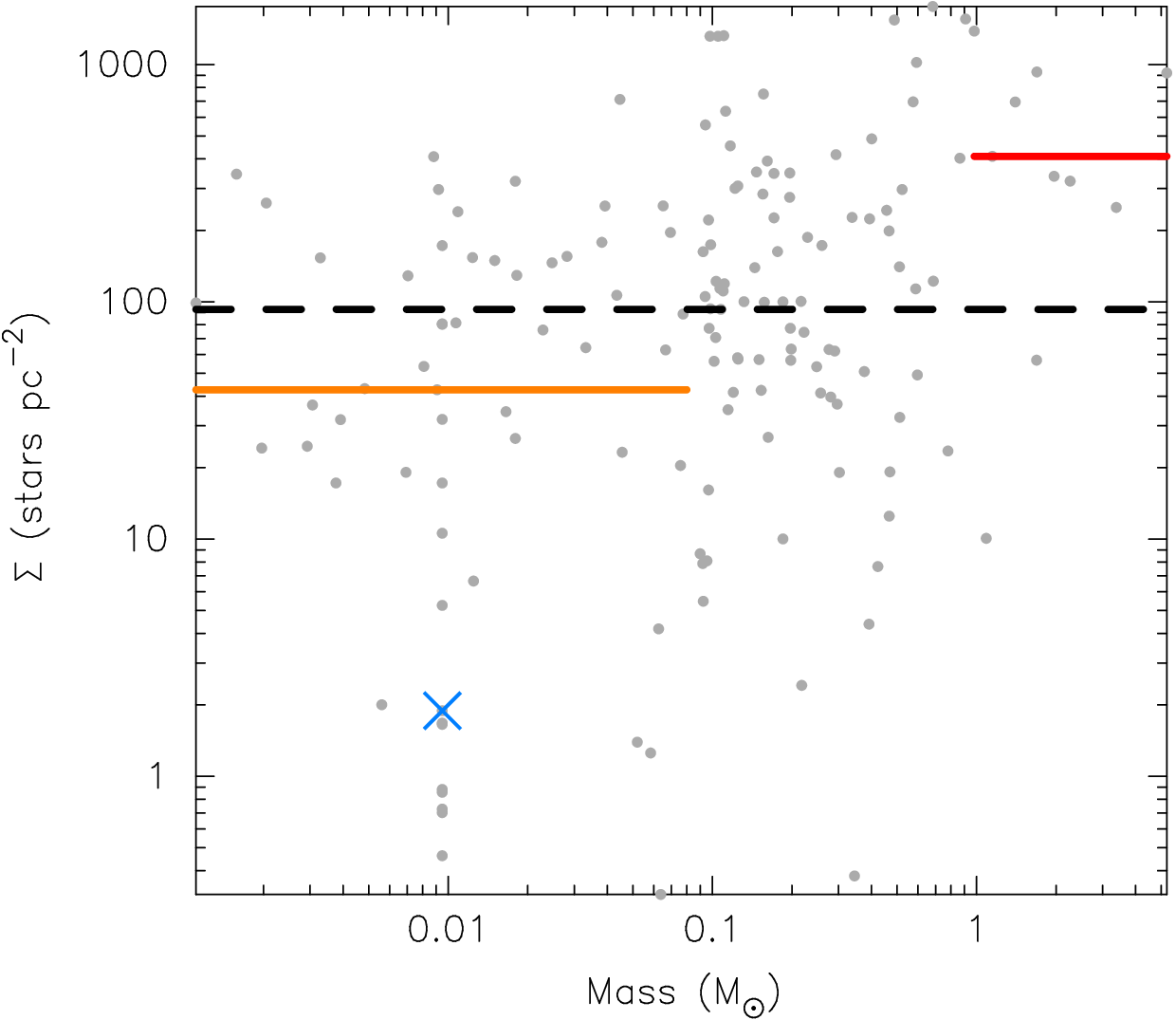}}
\end{center}
\caption[bf]{$\Sigma - m$ for one of our simulations. The surface density of each object, $\Sigma$ is plotted against its mass $m$. The median surface density for the star-forming region is shown by the dashed black line, the median surface density for the most massive stars is shown by the solid red line. The median surface density for brown dwarfs ($0.01 < m/{\rm M_\odot} < 0.08$) is shown by the solid orange line, and the median surface density for the ejected planetary mass objects (all of which are 10\,M$_{\rm Jup}$, i.e.\,\,$9.4 \times 10^{-2}$M$_\odot$) is shown by the blue cross. The planetary mass objects, which were ejected from stellar systems, have significantly lower local surface densities, whereas the brown dwarfs do not have significantly lower surface densities than the region as a whole.}
\label{sim_sigma-m}
\end{figure}

\begin{figure*}
  \begin{center}
\setlength{\subfigcapskip}{10pt}
\hspace*{-1.5cm} \subfigure[Proper motions]{\label{sim_velocities-a}\rotatebox{270}{\includegraphics[scale=0.32]{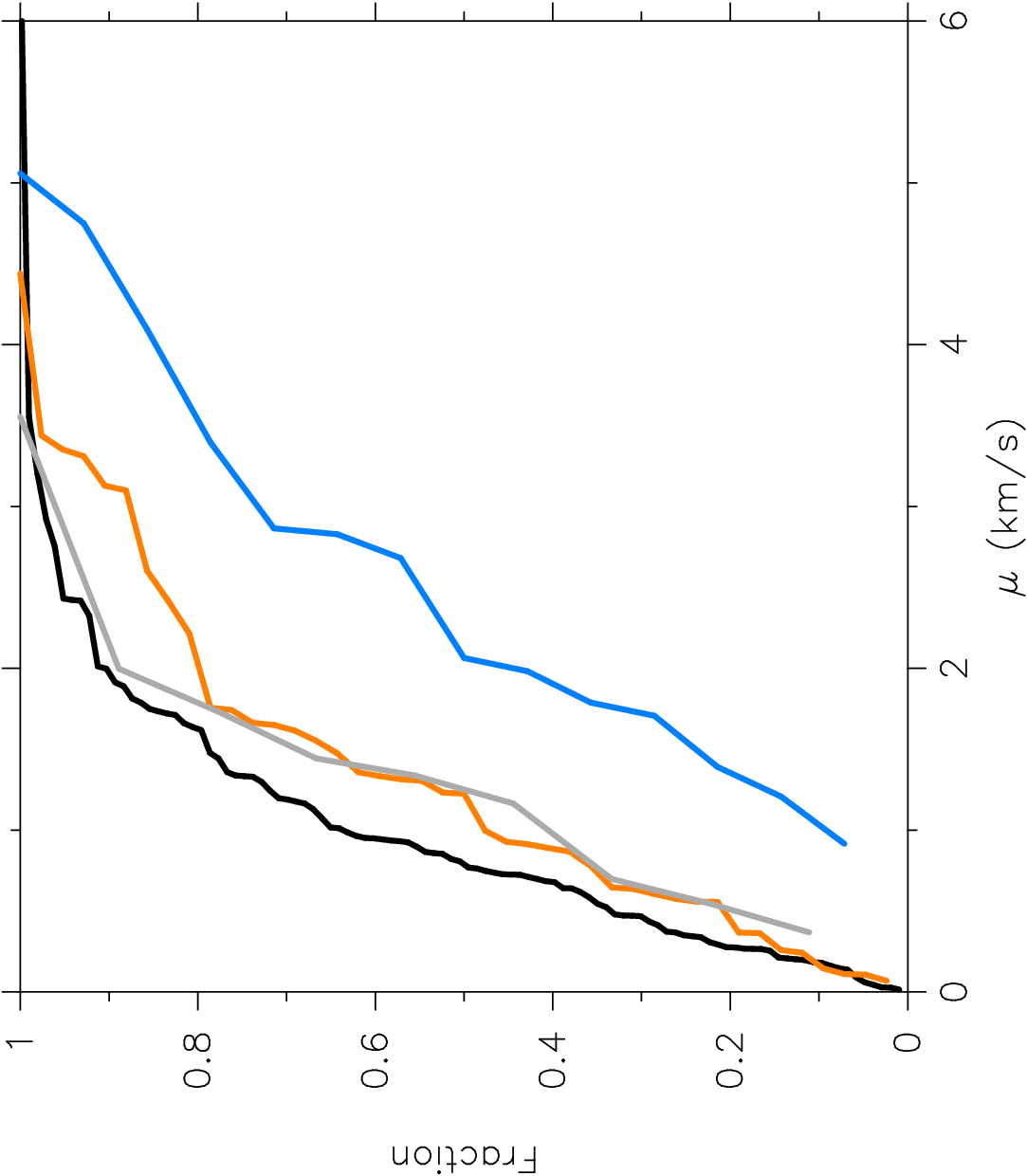}}} 
\hspace*{0.3cm}\subfigure[Radial velocities]{\label{sim_velocities-b}\rotatebox{270}{\includegraphics[scale=0.32]{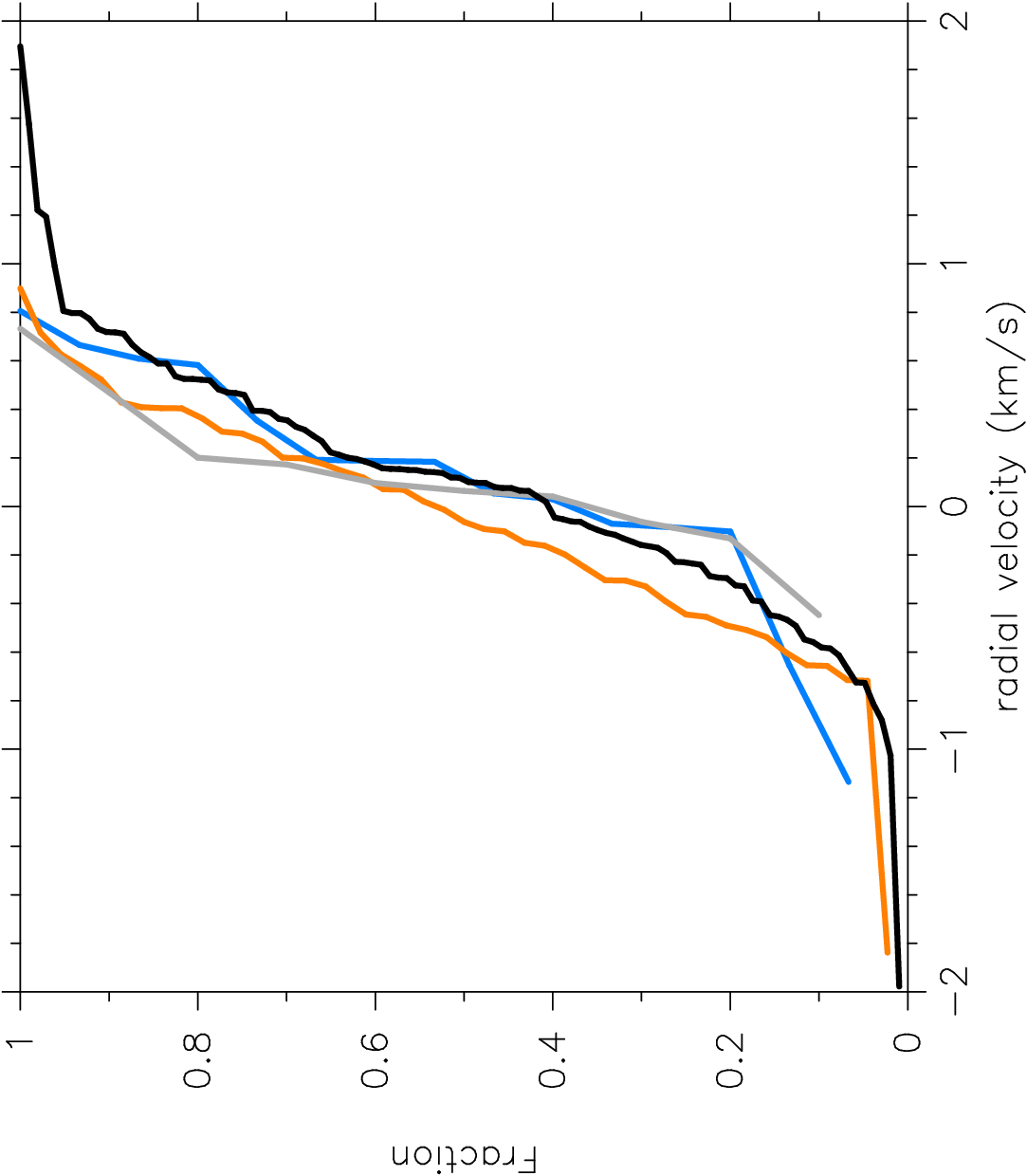}}}
\caption[bf]{Velocity distributions of stars, brown dwarfs and ejected planets in our simulations. The solid black lines are for all stars, the grey lines are stars that no longer host planets. The orange lines are the velocity distributions for brown dwarfs, and the blue lines are the ejected planets. Panel (a) shows the proper motion velocities, and panel (b) shows the radial velocities (the $v_z$ velocity component).}
\label{sim_velocities}
  \end{center}
\end{figure*}

\begin{figure}
\begin{center}
\rotatebox{270}{\includegraphics[scale=0.35]{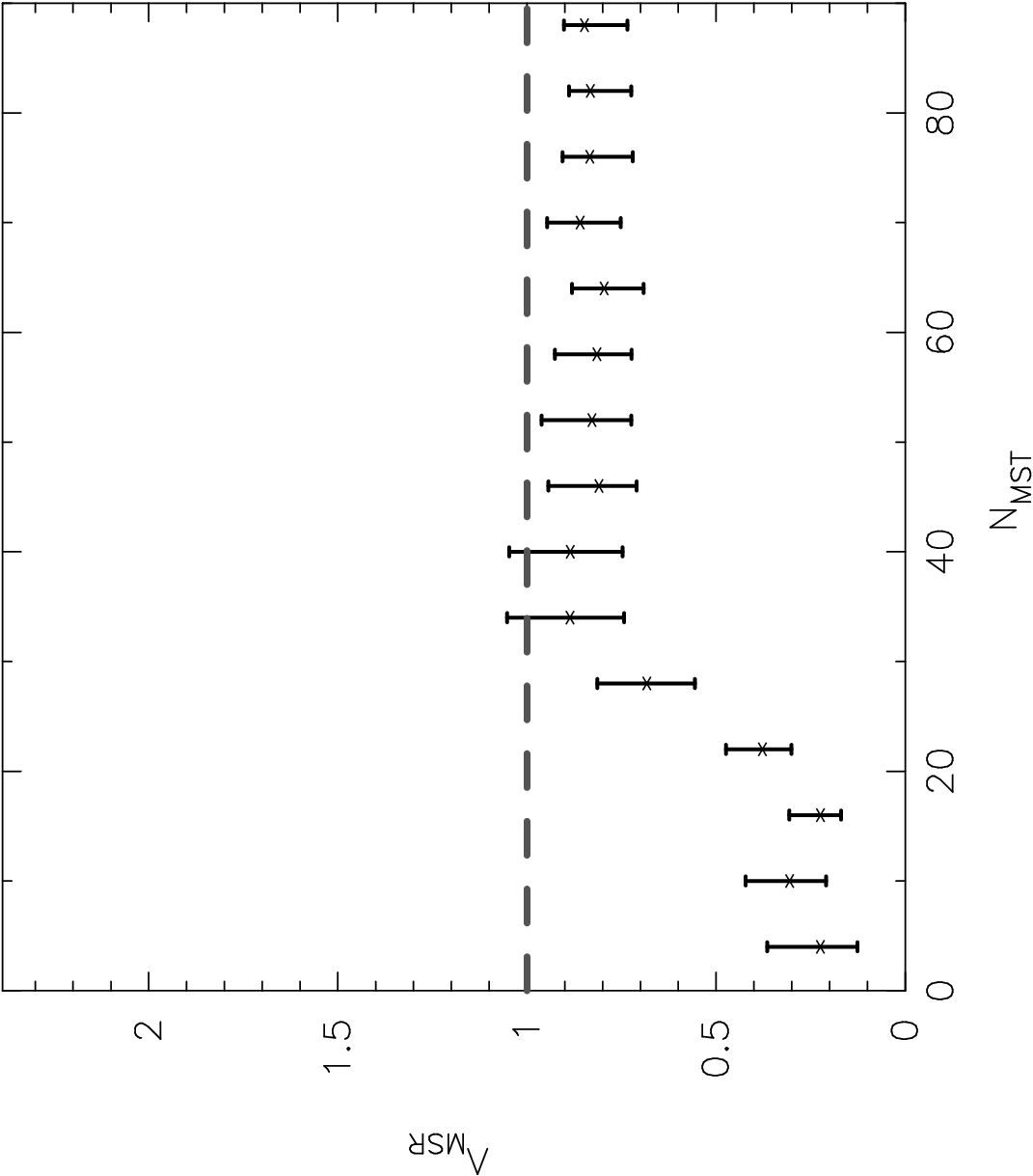}}
\end{center}
\caption[bf]{Calculation of the spatial distribution of brown dwarfs and planetary mass objects in our simulations with $\Lambda_{\rm MSR}$, where the planetary mass objects are less massive than the brown dwarfs (the planetary mass objects are all 1\,M$_{\rm Jup}$). We present the results when calculating $\Lambda_{\rm MSR}$ with the $N_{\rm MST} = 5$ least massive objects, and then subsequently adding the next 10 least massive objects. As the least massive objects are ejected, they are significantly more dispersed than the higher-mass brown dwarfs.}
\label{sim_JN_Lambda}
\end{figure}

\subsubsection{Mass segregation}

We calculate the mass segregation ratio, $\Lambda_{\rm MSR}$, using both the original method from \citet{Allison09a} and the slide method from \citet{Parker11b}. In the original determination of $\Lambda_{\rm MSR}$, we start with subsets of the lowest-mass objects (which are brown dwarfs) and add progressively more massive objects to the sample. This is shown in Fig.~\ref{sim_Lambda-a}. The least massive brown dwarfs are consistent with $\Lambda_{\rm MSR} = 1$, indicating no mass segregation, before $\Lambda_{\rm MSR} << 1$ for slightly more massive objects, which includes the planetary-mass objects.

We then calculate  $\Lambda_{\rm MSR}$ for discrete mass bins containing ten objects, starting from the least massive subset (Fig.~\ref{sim_Lambda-b}). The two least massive subsets contain brown dwarfs drawn from the mass function and with identical velocity and spatial distributions to the stars. These have $\Lambda_{\rm MSR}$ ratios consistent with unity.

The next subset (third bin from the left) contains planetary mass objects that were orbiting stars but have been ejected through dynamical encounters. They are significantly more widely distributed than the other objects in the star-forming region, with $\Lambda_{\rm MSR} = 0.30^{+0.41}_{-0.21}$. This is an extremely significant deviation from  $\Lambda_{\rm MSR} = 1$. At higher masses, but still within the brown dwarf regime, $\Lambda_{\rm MSR} \sim 1$.  The only other subset that deviates significantly from $\Lambda_{\rm MSR} = 1$ are the most massive stars (the rightmost bin in Fig.~\ref{sim_Lambda-a}).

\subsubsection{Relative surface densities}

In Fig.~\ref{sim_sigma-m} we plot the local surface density, $\Sigma$ of each object in the simulation snapshot, as a function of the mass of the object. We then compare bins of different types of objects. The median density of all objects in the star-forming region is shown by the dashed black line; this is 93\,stars\,pc$^{-2}$ at 3\,Myr. The median density of the brown dwarfs (all objects with mass $m < 0.08$\,M$_\odot$) is 43\,stars\,pc$^{-2}$, shown by the solid orange line. As in most of our simulations, the most massive stars have attained higher than average surface densities (400\,stars\,pc$^{-2}$, the solid red line). Finally, we show the median surface density of ejected planets by the blue cross (2\,stars\,pc$^{-2}$).

We assess the significance of the differences in densities between the subsets by using a KS test where we reject the null hypothesis that two subsets share the same underlying parent distribution if the $p$-value is less than 0.1.

The brown dwarfs do not have significantly lower densities than the star-forming region as a whole (a KS test returns a difference $D = 0.18$ with a $p-$value $p = 0.11$, and so we cannot reject the hypothesis that that the brown dwarfs share the same underlying density distribution as the stars). Note that the brown dwarf subset contains the ejected planetary-mass objects. If we take the ejected planets as their own subset, we find that these do have significantly lower densities than all of the objects in the star-forming region, with a KS test returning a difference $D = 0.46$ with an associated $p$-value $p = 2 \times 10^{-3}$.\\

Whilst measuring different things, both $\Lambda_{\rm MSR}$ and $\Sigma - m$ show that substellar objects that formed around stars but were then ejected are likely to have a significantly different spatial distribution to substellar objects that formed like the stars in the region.

\subsubsection{Velocity distributions}

We now compare the velocity distributions of the brown dwarfs and ejected planets to the stars. In Fig.~\ref{sim_velocities-a} we show the proper motions of the ejected planets (the blue line), the brown dwarfs (the orange line), single stars (i.e.\,\,stars without a planetary companion, the grey line) and all stars (the black line). The free-floating planets clearly are moving at faster velocities (likely to be as a result of their ejection), whereas the brown dwarfs are moving at similar (albeit slightly faster) velocities compared to the stars.

Conversely, the radial velocity distributions (Fig.~\ref{sim_velocities-b}) are quite similar. The velocity dispersion for the ejected planets is 0.49\,km\,s$^{-1}$ (the blue line), whereas for single stars (grey line) it is 0.30\,km\,s$^{-1}$. For all stars, the dispersion is larger (0.54\,km\,s$^{-1}$, the black line), but this is likely to be inflated by the contribution from the planetary companions still orbiting the majority of stars \citep{Gieles10,Cottaar12b}.

\subsubsection{Lower planetary masses}

We repeat the above analysis, but this time for identical simulations, save for the planet masses, which are now 1\,M$_{\rm Jup}$, i.e.\,\,$9.4 \times 10^{-4}$\,M$_\odot$. This means that the planetary mass objects are slightly lower mass than the brown dwarfs that formed like stars in the simulation.

In Fig.~\ref{sim_JN_Lambda} we see that the signature of inverse mass segregation of the ejected planetary-mass objects is more obvious than when the planets have masses similar to (or higher than some of) the brown dwarfs. These simulations also display similar behaviour to the previous models in both the $\Lambda_{\rm MSR}$ slide and $\Sigma - m$ plots.  There are 16 free-floating planetary mass objects, more spatially distributed than the brown dwarfs, and Fig.~\ref{sim_JN_Lambda} shows that even when the bins are not independent of one another, the difference in the $\Lambda_{\rm MSR}$ measurement disappears once a further 15--20 brown dwarfs are included in the calculation.

Lowering the planetary companion masses has the effect of lowering the binding energy of the system, which theoretically makes the system more susceptible to destruction. However, \citet{Parker13b} show for stellar binaries that the companion mass ratio has very little influence on whether the system will be destroyed, because the typical interaction that breaks apart a system has an energy more than ten times that of the binding energy of the system. We therefore do not expect to see a significant difference in the kinematic distributions of the ejected 1\,M$_{\rm Jup}$ planets compared to 10\,M$_{\rm Jup}$ and this is the case for our simulations.

\subsubsection{Closer planetary orbits}

In one set of simulations, we placed the planets at 5\,au around their host stars, rather than 30\,au. This difference reduces the number of planets that are liberated from their host stars  (because the systems are dynamically `harder' according to the Heggie-Hills law, \citealp{Heggie75,Hills75a}) by a factor of two \citep[similar to results previously reported by][]{Parker12a}, but we find that the spatial and kinematic distributions of these ejected planets are the same as in those simulations where the planets are originally 30\,au from their host stars.

\subsubsection{Lower stellar densities, less substructure}

In the simulations where the stellar density is lower, we produce fewer free-floating planets, but again, the planets that are liberated from their host stars have a different spatial and kinematic distribution to the brown dwarfs.

\section{Discussion}

We find no evidence that brown dwarfs or planetary mass objects have a different spatial distribution to the stars in NGC\,1333. However, in $N$-body simulations we show that a population of planetary-mass objects created via ejection from a bound orbit around a star would have a significantly different spatial distribution to the stars, and also any brown dwarfs that formed like stars, representing a continuous extension to the low-mass end of the IMF.

Our results should be prefaced by several caveats.

First, we may not be accurately representing the initial conditions of NGC\,1333. In a previous study \citep{Parker17a}, we showed that NGC\,1333 was likely initially quite dense, but in many of the simulations we use here, clear mass segregation of the most massive stars occurs, which is not observed in NGC\,1333. Therefore, our simulations may be too dense, although reducing the initial stellar density would merely reduce the number of free-floating planetary-mass objects, thus unlikely to affect our conclusions.

Second, the simulations -- by  definition -- assume instantaneous star and planet formation. In reality, even the shortest estimates suggest star formation takes up to 1\,Myr \citep{Elmegreen00}, and gas giants are likely to take just as long to form \citep{Alves20,SeguraCox20}. However, what we sacrifice in realism in the simulations is compensated for in the statistical significance we gain from running multiple versions of the same simulations.

Third, our planetary `systems' consist of just one planet, placed at either 5\,au (Jupiter's location in the Solar System) or 30\,au (Neptune's location). Our fractal simulations are unable to accommodate mult-planet systems, although this is likely to change in the near future. Therefore, our simulations cannot account for planet--planet interactions once the outer orbiting planet has been destabilised by an interaction with a star \citep{Malmberg07a}. Further planet--planet interactions are more likely to produce even more free-floating planets with different spatial distributions to the brown dwarfs.

Additionally, our choice of initial semimajor axes and other orbital parameters (zero eccentricity/inclination) could affect the spatial and velocity distribution of the free-floating planets. In practice, our chosen initial orbital parameters likely straddle the median values for the semimajor axis distributions \citep{Forgan15,Zheng15}, and as such the population of free-floating planets represents the average outcome of dynamical encounters in star-forming regions \citep{DaffernPowell22}.

Finally, our simulations do not contain primordial stellar/brown dwarf binaries (with the only binary systems being the star--planet systems). Binary systems would slightly increase the number of destructive encounters due to the slightly higher collisional cross section. If binary companions were brown dwarfs, then systems broken up from encounters could produce a population of brown dwarfs with a similar spatial distribution to the free-floating planets. The binary fraction for brown dwarf--brown dwarf systems is quite low \citep[$\sim 15$\,per  cent,][]{Burgasser07}, but brown dwarfs could be companions to M-dwarfs, which have a higher binary fraction \citep[up to 30\,per cent,][]{Ward-Duong15}. We will investigate the effects of binarity on mass segregation more generally in a future paper.

We also reiterate that mass segregation (and inverse mass segregation) can be a very transient phenomenon, in that it can appear, then disappear, then reappear. This often happens when stars are ejected, and in several of our simulations we see inverse mass segregation of the planetary mass objects at e.g. 3\,Myr, but not later after the PMOs have been ejected \citep[and hence discounted from the analysis as an observer would only be able to trace them back to the origin with e.g. Gaia,][]{Schoettler20}. Whilst the PMOs appear more spatially distributed than stars and brown dwarfs in the majority of our simulation snapshots, we cannot fully rule out the possibility that the PMOs in NGC\,1333 were previously more spread out, or will become more spread out at later times.

Despite this, our analysis of the low-mass objects in NGC\,1333 indicates that these objects follow the spatial distribution of the stars, whereas our simulations generally show that the planetary mass objects that were ejected from stellar systems would be more spread out, and also moving with faster proper motion velocities, than stars and substellar objects that formed like stars, i.e.\,\,from the collapse and fragmentation of the host GMC.

This is even evident in the simulations in which the planets that are ejected from orbits around stars have a mass that overlaps with the brown dwarf mass regime. The planetary-mass objects that are ejected always have a very different spatial and kinematic distribution to those that formed more like stars, irrespective of their initial mass or semimajor axis.

  Aside from these spatial and kinematic signatures, there is not a clear diagnostic that can be used to distinguish between brown dwarf objects that formed like stars and objects that formed like planets, but a more accurate determination of their masses could help identify their formation mechanism.

  For solar metallicity GMCs, 1\,M$_{\rm Jup}$ planets are much lower than the opacity limit for fragmentation \citep{Rees76,Whitworth06,Bate14} and so would probably be ejected planets \citep[though see e.g.][]{Boss01}. On the other hand, forming $>$10\,M$_{\rm Jup}$ planets by core accretion in a circumstellar disc is likely to be challenging \citep{BergezCasalou23,Helled23}, and these objects (which encompass the PMOs found in NGC\,1333 to date) probably form more like stars, although some could form via disc fragmentation \citep[e.g.][]{Mayer02,Stamatellos09}, depending on the physical conditions in the disc \citep{Meru12,Kratter16}.

The PMOs observed by \citet{Scholz23} in NGC\,1333 are therefore likely to be the tail of the initial mass function, and formed in the same way as stars, rather than being the result of ejections from planetary systems. Of course, there may be low- and planetary-mass objects in NGC\,1333 that have not yet been discovered, and we would expect these to be preferentially found on the outskirts (where the observational completeness is lower). We also note that whilst dynamical encounters produce around 10 free-floating planets in our simulations, these planets are at relatively large distances from their host stars (30\,au). For planets on smaller orbits (e.g. 5\,au) the number of free-floating planets produced through dynamical encounters is reduced by a factor of two.

Although none of the PMOs in NGC\,1333 appear to have formed as planets and then been ejected form their host star, \citet{Osorio14b} find evidence that planetary-mass objects in the Pleiades open cluster appear to be moving with faster proper motions than the stars. As star clusters never reach complete energy equipartition \citep{Spitzer69,Trenti13,Parker16c,Spera16}, these objects are likely to be ejected planets that are now free-floating in the cluster, rather than objects that formed like stars that have susequently attained higher velocities due to repeated interactions. 

We therefore encourage further observational studies of the substellar population in star-forming regions to both characterise the substellar population and to determine whether planetary-mass objects could be the result of dynamical encounters.

\section{Conclusions}

We have quantified the spatial distribution of the substellar population of NGC\,1333 to determine whether planetary-mass objects have a different spatial distribution to stars. We then analyse $N$-body simulations containing both brown dwarfs that form in the same way as stars, and high-mass planets originally orbiting stars, to compare their respective spatial distributions. Our conclusions are the following.

(i) The brown dwarfs and planetary mass objects in NGC\,1333 follow the same spatial distribution as the stars according to the $\Lambda_{\rm MSR}$ mass segregation ratio, and the relative surface density metric $\Sigma - m$.

(ii) In $N$-body simulations planets are liberated from their host stars and form a spatially distinct population from the brown dwarfs, which were set up to form in the same way as stars.

(iii) The difference between these populations can still be discerned even if the mass ranges overlap, i.e.\,\,if the planets have a higher mass than some of the brown dwarfs.

(iv) The substellar objects observed in NGC\,1333 are therefore unlikely to be free-floating planets created as a result of dynamical interactions, having previously orbited stars. Rather, the PMOs in NGC\,1333 likely formed in a similar way to the stellar, and higher-mass substellar populations. 
 
(v) If there is a population of ejected free-floating planets in NGC\,1333 (or in other star-forming regions), we would expect to observe these objects on the outskirts of the region, where current observations are likely incomplete. As such, observations with e.g.\,\,JWST NIRSpec will be crucial to untangling the substellar populations in star-forming regions.

\section*{Data availability statement}

The data used to produce the plots in this paper will be shared on reasonable request to the corresponding author.

\section*{Acknowledgments}

We thank the anonymous referee for a helpful report. RJP acknowledges support from the Royal Society in the form of a Dorothy Hodgkin Fellowship.

\bibliographystyle{mnras}
\bibliography{general_ref}

\begin{thebibliography}{}
\makeatletter
\relax
\def\mn@urlcharsother{\let\do\@makeother \do\$\do\&\do\#\do\^\do\_\do\%\do\~}
\def\mn@doi{\begingroup\mn@urlcharsother \@ifnextchar [ {\mn@doi@}
  {\mn@doi@[]}}
\def\mn@doi@[#1]#2{\def\@tempa{#1}\ifx\@tempa\@empty \href
  {http://dx.doi.org/#2} {doi:#2}\else \href {http://dx.doi.org/#2} {#1}\fi
  \endgroup}
\def\mn@eprint#1#2{\mn@eprint@#1:#2::\@nil}
\def\mn@eprint@arXiv#1{\href {http://arxiv.org/abs/#1} {{\tt arXiv:#1}}}
\def\mn@eprint@dblp#1{\href {http://dblp.uni-trier.de/rec/bibtex/#1.xml}
  {dblp:#1}}
\def\mn@eprint@#1:#2:#3:#4\@nil{\def\@tempa {#1}\def\@tempb {#2}\def\@tempc
  {#3}\ifx \@tempc \@empty \let \@tempc \@tempb \let \@tempb \@tempa \fi \ifx
  \@tempb \@empty \def\@tempb {arXiv}\fi \@ifundefined
  {mn@eprint@\@tempb}{\@tempb:\@tempc}{\expandafter \expandafter \csname
  mn@eprint@\@tempb\endcsname \expandafter{\@tempc}}}

\bibitem[\protect\citeauthoryear{{Alfaro} \& {Gonz{\'a}lez}}{{Alfaro} \&
  {Gonz{\'a}lez}}{2016}]{Alfaro16}
{Alfaro} E.~J.,  {Gonz{\'a}lez} M.,  2016, \mn@doi [MNRAS]
  {10.1093/mnras/stv2822}, \href
  {http://adsabs.harvard.edu/abs/2016MNRAS.456.2900A} {456, 2900}

\bibitem[\protect\citeauthoryear{{Alfaro} \&
  {Rom{\'a}n-Z{\'u}{\~n}iga}}{{Alfaro} \&
  {Rom{\'a}n-Z{\'u}{\~n}iga}}{2018}]{Alfaro18}
{Alfaro} E.~J.,  {Rom{\'a}n-Z{\'u}{\~n}iga} C.~G.,  2018, \mn@doi [\mnras]
  {10.1093/mnrasl/sly075}, \href
  {http://adsabs.harvard.edu/abs/2018MNRAS.478L.110A} {478, L110}

\bibitem[\protect\citeauthoryear{Allison, Goodwin, Parker, {Portegies Zwart},
  de Grijs  \& Kouwenhoven}{Allison et~al.}{2009}]{Allison09a}
Allison R.~J.,  Goodwin S.~P.,  Parker R.~J.,  {Portegies Zwart} S.~F.,  de
  Grijs R.,   Kouwenhoven M. B.~N.,  2009, MNRAS, 395, 1449

\bibitem[\protect\citeauthoryear{{Alves}, {Cleeves}, {Girart}, {Zhu}, {Franco},
  {Zurlo}  \& {Caselli}}{{Alves} et~al.}{2020}]{Alves20}
{Alves} F.~O.,  {Cleeves} L.~I.,  {Girart} J.~M.,  {Zhu} Z.,  {Franco} G.
  A.~P.,  {Zurlo} A.,   {Caselli} P.,  2020, \mn@doi [\apjl]
  {10.3847/2041-8213/abc550}, \href
  {https://ui.adsabs.harvard.edu/abs/2020ApJ...904L...6A} {904, L6}

\bibitem[\protect\citeauthoryear{Andersen, Meyer, Greissl  \& Aversa}{Andersen
  et~al.}{2008}]{Andersen08}
Andersen M.,  Meyer M.~R.,  Greissl J.,   Aversa A.,  2008, ApJL, 683, L183

\bibitem[\protect\citeauthoryear{{Baraffe}, {Chabrier}, {Allard}  \&
  {Hauschildt}}{{Baraffe} et~al.}{2002}]{Baraffe02}
{Baraffe} I.,  {Chabrier} G.,  {Allard} F.,   {Hauschildt} P.~H.,  2002,
  \mn@doi [\aap] {10.1051/0004-6361:20011638}, \href
  {https://ui.adsabs.harvard.edu/abs/2002A&A...382..563B} {382, 563}

\bibitem[\protect\citeauthoryear{{Barrado y Navascu{\'e}s}, {Bouvier},
  {Stauffer}, {Lodieu}  \& {McCaughrean}}{{Barrado y Navascu{\'e}s}
  et~al.}{2002}]{Barrado02}
{Barrado y Navascu{\'e}s} D.,  {Bouvier} J.,  {Stauffer} J.~R.,  {Lodieu} N.,
  {McCaughrean} M.~J.,  2002, \mn@doi [\aap] {10.1051/0004-6361:20021262},
  \href {https://ui.adsabs.harvard.edu/abs/2002A&A...395..813B} {395, 813}

\bibitem[\protect\citeauthoryear{Bastian, Covey  \& Meyer}{Bastian
  et~al.}{2010}]{Bastian10}
Bastian N.,  Covey K.~R.,   Meyer M.~R.,  2010, ARA\&A, 48, 339

\bibitem[\protect\citeauthoryear{Bate}{Bate}{2009}]{Bate09}
Bate M.~R.,  2009, MNRAS, 392, 590

\bibitem[\protect\citeauthoryear{{Bate}}{{Bate}}{2014}]{Bate14}
{Bate} M.~R.,  2014, \mn@doi [\mnras] {10.1093/mnras/stu795}, \href
  {https://ui.adsabs.harvard.edu/abs/2014MNRAS.442..285B} {442, 285}

\bibitem[\protect\citeauthoryear{{Bergez-Casalou}, {Bitsch}  \&
  {Raymond}}{{Bergez-Casalou} et~al.}{2023}]{BergezCasalou23}
{Bergez-Casalou} C.,  {Bitsch} B.,   {Raymond} S.~N.,  2023, \mn@doi [\aap]
  {10.1051/0004-6361/202244988}, \href
  {https://ui.adsabs.harvard.edu/abs/2023A&A...669A.129B} {669, A129}

\bibitem[\protect\citeauthoryear{{Blaylock-Squibbs}, {Parker}, {Buckner}  \&
  {G{\"u}del}}{{Blaylock-Squibbs} et~al.}{2022}]{BlaylockSquibbs22}
{Blaylock-Squibbs} G.~A.,  {Parker} R.~J.,  {Buckner} A. S.~M.,   {G{\"u}del}
  M.,  2022, \mn@doi [\mnras] {10.1093/mnras/stab3447}, \href
  {https://ui.adsabs.harvard.edu/abs/2022MNRAS.510.2864B} {510, 2864}

\bibitem[\protect\citeauthoryear{{Boley}, {Payne}  \& {Ford}}{{Boley}
  et~al.}{2012}]{Boley12}
{Boley} A.~C.,  {Payne} M.~J.,   {Ford} E.~B.,  2012, \mn@doi [\apj]
  {10.1088/0004-637X/754/1/57}, \href
  {https://ui.adsabs.harvard.edu/abs/2012ApJ...754...57B} {754, 57}

\bibitem[\protect\citeauthoryear{Bonnell, Smith, Davies  \& Horne}{Bonnell
  et~al.}{2001}]{Bonnell01b}
Bonnell I.~A.,  Smith K.~W.,  Davies M.~B.,   Horne K.,  2001, MNRAS, 322, 859

\bibitem[\protect\citeauthoryear{{Boss}}{{Boss}}{2001}]{Boss01}
{Boss} A.~P.,  2001, \mn@doi [\apjl] {10.1086/320033}, \href
  {https://ui.adsabs.harvard.edu/abs/2001ApJ...551L.167B} {551, L167}

\bibitem[\protect\citeauthoryear{{Buckner} et~al.,}{{Buckner}
  et~al.}{2019}]{Buckner19}
{Buckner} A. S.~M.,  et~al., 2019, \mn@doi [\aap]
  {10.1051/0004-6361/201832936}, \href
  {https://ui.adsabs.harvard.edu/abs/2019A&A...622A.184B} {622, A184}

\bibitem[\protect\citeauthoryear{Burgasser, Reid, Siegler, Close, Allen,
  Lowrance  \& Gizis}{Burgasser et~al.}{2007}]{Burgasser07}
Burgasser A.~J.,  Reid I.~N.,  Siegler N.,  Close L.,  Allen P.,  Lowrance P.,
   Gizis J.,  2007, in Reipurth B.,  Jewitt D.,   Keil K.,  eds, {Protostars
  and Planets V}. pp 427--441

\bibitem[\protect\citeauthoryear{{Canty}, {Lucas}, {Roche}  \&
  {Pinfield}}{{Canty} et~al.}{2013}]{Canty13}
{Canty} J.~I.,  {Lucas} P.~W.,  {Roche} P.~F.,   {Pinfield} D.~J.,  2013,
  \mn@doi [\mnras] {10.1093/mnras/stt1477}, \href
  {https://ui.adsabs.harvard.edu/abs/2013MNRAS.435.2650C} {435, 2650}

\bibitem[\protect\citeauthoryear{Cartwright \& Whitworth}{Cartwright \&
  Whitworth}{2004}]{Cartwright04}
Cartwright A.,  Whitworth A.~P.,  2004, MNRAS, 348, 589

\bibitem[\protect\citeauthoryear{Casertano \& Hut}{Casertano \&
  Hut}{1985}]{Casertano85}
Casertano S.,  Hut P.,  1985, ApJ, 298, 80

\bibitem[\protect\citeauthoryear{{Chabrier}, {Johansen}, {Janson}  \&
  {Rafikov}}{{Chabrier} et~al.}{2014}]{Chabrier14}
{Chabrier} G.,  {Johansen} A.,  {Janson} M.,   {Rafikov} R.,  2014, in
  {Beuther} H.,  {Klessen} R.~S.,  {Dullemond} C.~P.,   {Henning} T.,  eds,
  Protostars and Planets VI. pp 619--642 (\mn@eprint {arXiv} {1401.7559}),
  \mn@doi{10.2458/azu_uapress_9780816531240-ch027}

\bibitem[\protect\citeauthoryear{{Chatterjee}, {Ford}, {Matsumura}  \&
  {Rasio}}{{Chatterjee} et~al.}{2008}]{Chatterjee08}
{Chatterjee} S.,  {Ford} E.~B.,  {Matsumura} S.,   {Rasio} F.~A.,  2008,
  \mn@doi [\apj] {10.1086/590227}, \href
  {https://ui.adsabs.harvard.edu/abs/2008ApJ...686..580C} {686, 580}

\bibitem[\protect\citeauthoryear{{Cottaar}, {Meyer}  \& {Parker}}{{Cottaar}
  et~al.}{2012}]{Cottaar12b}
{Cottaar} M.,  {Meyer} M.~R.,   {Parker} R.~J.,  2012, \mn@doi [A\&A]
  {10.1051/0004-6361/201219673}, 547, A35

\bibitem[\protect\citeauthoryear{{Daffern-Powell} \& {Parker}}{{Daffern-Powell}
  \& {Parker}}{2020}]{DaffernPowell20}
{Daffern-Powell} E.~C.,  {Parker} R.~J.,  2020, \mn@doi [\mnras]
  {10.1093/mnras/staa575}, \href
  {https://ui.adsabs.harvard.edu/abs/2020MNRAS.493.4925D} {493, 4925}

\bibitem[\protect\citeauthoryear{{Daffern-Powell}, {Parker}  \&
  {Quanz}}{{Daffern-Powell} et~al.}{2022}]{DaffernPowell22}
{Daffern-Powell} E.~C.,  {Parker} R.~J.,   {Quanz} S.~P.,  2022, \mn@doi
  [\mnras] {10.1093/mnras/stac1392}, \href
  {https://ui.adsabs.harvard.edu/abs/2022MNRAS.514..920D} {514, 920}

\bibitem[\protect\citeauthoryear{Elmegreen}{Elmegreen}{2000}]{Elmegreen00}
Elmegreen B.~G.,  2000, ApJ, 530, 277

\bibitem[\protect\citeauthoryear{{Esplin} \& {Luhman}}{{Esplin} \&
  {Luhman}}{2017}]{Esplin17}
{Esplin} T.~L.,  {Luhman} K.~L.,  2017, \mn@doi [\aj]
  {10.3847/1538-3881/aa859b}, \href
  {https://ui.adsabs.harvard.edu/abs/2017AJ....154..134E} {154, 134}

\bibitem[\protect\citeauthoryear{{Feiden} \& {Chaboyer}}{{Feiden} \&
  {Chaboyer}}{2012}]{Feiden12}
{Feiden} G.~A.,  {Chaboyer} B.,  2012, \mn@doi [\apj]
  {10.1088/0004-637X/757/1/42}, \href
  {https://ui.adsabs.harvard.edu/abs/2012ApJ...757...42F} {757, 42}

\bibitem[\protect\citeauthoryear{{Forgan}, {Parker}  \& {Rice}}{{Forgan}
  et~al.}{2015}]{Forgan15}
{Forgan} D.,  {Parker} R.~J.,   {Rice} K.,  2015, \mn@doi [\mnras]
  {10.1093/mnras/stu2504}, \href
  {https://ui.adsabs.harvard.edu/abs/2015MNRAS.447..836F} {447, 836}

\bibitem[\protect\citeauthoryear{{Gagn{\'e}} et~al.,}{{Gagn{\'e}}
  et~al.}{2017}]{Gagne17}
{Gagn{\'e}} J.,  et~al., 2017, \mn@doi [\apjs] {10.3847/1538-4365/228/2/18},
  \href {https://ui.adsabs.harvard.edu/abs/2017ApJS..228...18G} {228, 18}

\bibitem[\protect\citeauthoryear{{Gahm}, {Grenman}, {Fredriksson}  \&
  {Kristen}}{{Gahm} et~al.}{2007}]{Gahm07}
{Gahm} G.~F.,  {Grenman} T.,  {Fredriksson} S.,   {Kristen} H.,  2007, \mn@doi
  [\aj] {10.1086/512036}, \href
  {https://ui.adsabs.harvard.edu/abs/2007AJ....133.1795G} {133, 1795}

\bibitem[\protect\citeauthoryear{{Geers}, {Scholz}, {Jayawardhana}, {Lee},
  {Lafreni{\`e}re}  \& {Tamura}}{{Geers} et~al.}{2011}]{Geers11}
{Geers} V.,  {Scholz} A.,  {Jayawardhana} R.,  {Lee} E.,  {Lafreni{\`e}re} D.,
   {Tamura} M.,  2011, \mn@doi [\apj] {10.1088/0004-637X/726/1/23}, \href
  {https://ui.adsabs.harvard.edu/abs/2011ApJ...726...23G} {726, 23}

\bibitem[\protect\citeauthoryear{{Gieles}, {Sana}  \& {Portegies
  Zwart}}{{Gieles} et~al.}{2010}]{Gieles10}
{Gieles} M.,  {Sana} H.,   {Portegies Zwart} S.~F.,  2010, \mn@doi [MNRAS]
  {10.1111/j.1365-2966.2009.15993.x}, \href
  {http://adsabs.harvard.edu/abs/2010MNRAS.402.1750G} {402, 1750}

\bibitem[\protect\citeauthoryear{{Girichidis}, {Federrath}, {Allison},
  {Banerjee}  \& {Klessen}}{{Girichidis} et~al.}{2012}]{Girichidis12}
{Girichidis} P.,  {Federrath} C.,  {Allison} R.,  {Banerjee} R.,   {Klessen}
  R.~S.,  2012, MNRAS, 420, 3264

\bibitem[\protect\citeauthoryear{{Gonz{\'a}lez} \& {Alfaro}}{{Gonz{\'a}lez} \&
  {Alfaro}}{2017}]{Gonzalez17}
{Gonz{\'a}lez} M.,  {Alfaro} E.~J.,  2017, \mn@doi [\mnras]
  {10.1093/mnras/stw2855}, \href
  {http://adsabs.harvard.edu/abs/2017MNRAS.465.1889G} {465, 1889}

\bibitem[\protect\citeauthoryear{Goodwin \& Whitworth}{Goodwin \&
  Whitworth}{2004}]{Goodwin04a}
Goodwin S.~P.,  Whitworth A.~P.,  2004, A\&A, 413, 929

\bibitem[\protect\citeauthoryear{{Guszejnov}, {Markey}, {Offner}, {Grudi{\'c}},
  {Faucher-Gigu{\`e}re}, {Rosen}  \& {Hopkins}}{{Guszejnov}
  et~al.}{2022}]{Guszejnov22}
{Guszejnov} D.,  {Markey} C.,  {Offner} S. S.~R.,  {Grudi{\'c}} M.~Y.,
  {Faucher-Gigu{\`e}re} C.-A.,  {Rosen} A.~L.,   {Hopkins} P.~F.,  2022,
  \mn@doi [\mnras] {10.1093/mnras/stac1737}, \href
  {https://ui.adsabs.harvard.edu/abs/2022MNRAS.515..167G} {515, 167}

\bibitem[\protect\citeauthoryear{{Hacar}, {Tafalla}, {Kauffmann}  \&
  {Kov{\'a}cs}}{{Hacar} et~al.}{2013}]{Hacar13}
{Hacar} A.,  {Tafalla} M.,  {Kauffmann} J.,   {Kov{\'a}cs} A.,  2013, A\&A,
  554, A55

\bibitem[\protect\citeauthoryear{{Haworth}, {Facchini}  \& {Clarke}}{{Haworth}
  et~al.}{2015}]{Haworth15}
{Haworth} T.~J.,  {Facchini} S.,   {Clarke} C.~J.,  2015, \mn@doi [\mnras]
  {10.1093/mnras/stu2174}, \href
  {https://ui.adsabs.harvard.edu/abs/2015MNRAS.446.1098H} {446, 1098}

\bibitem[\protect\citeauthoryear{Heggie}{Heggie}{1975}]{Heggie75}
Heggie D.~C.,  1975, MNRAS, 173, 729

\bibitem[\protect\citeauthoryear{{Helled}}{{Helled}}{2023}]{Helled23}
{Helled} R.,  2023, \mn@doi [\aap] {10.1051/0004-6361/202346850}, \href
  {https://ui.adsabs.harvard.edu/abs/2023A&A...675L...8H} {675, L8}

\bibitem[\protect\citeauthoryear{{Hetem} \& {Gregorio-Hetem}}{{Hetem} \&
  {Gregorio-Hetem}}{2019}]{Hetem19}
{Hetem} A.,  {Gregorio-Hetem} J.,  2019, \mn@doi [\mnras]
  {10.1093/mnras/stz2698}, \href
  {https://ui.adsabs.harvard.edu/abs/2019MNRAS.490.2521H} {490, 2521}

\bibitem[\protect\citeauthoryear{Hills}{Hills}{1975}]{Hills75a}
Hills J.~G.,  1975, AJ, 80, 809

\bibitem[\protect\citeauthoryear{{Jakobsen} et~al.,}{{Jakobsen}
  et~al.}{2022}]{Jakobsen22}
{Jakobsen} P.,  et~al., 2022, \mn@doi [\aap] {10.1051/0004-6361/202142663},
  \href {https://ui.adsabs.harvard.edu/abs/2022A&A...661A..80J} {661, A80}

\bibitem[\protect\citeauthoryear{{K{\"o}nyves} et~al.,}{{K{\"o}nyves}
  et~al.}{2020}]{Konyves20}
{K{\"o}nyves} V.,  et~al., 2020, \mn@doi [\aap] {10.1051/0004-6361/201834753},
  \href {https://ui.adsabs.harvard.edu/abs/2020A&A...635A..34K} {635, A34}

\bibitem[\protect\citeauthoryear{{Kratter} \& {Lodato}}{{Kratter} \&
  {Lodato}}{2016}]{Kratter16}
{Kratter} K.,  {Lodato} G.,  2016, \mn@doi [\araa]
  {10.1146/annurev-astro-081915-023307}, \href
  {https://ui.adsabs.harvard.edu/abs/2016ARA&A..54..271K} {54, 271}

\bibitem[\protect\citeauthoryear{Kraus \& Hillenbrand}{Kraus \&
  Hillenbrand}{2008}]{Kraus08}
Kraus A.~L.,  Hillenbrand L.~A.,  2008, ApJ, 686, L111

\bibitem[\protect\citeauthoryear{{Kubiak}, {Mu{\v{z}}i{\'c}}, {Sousa},
  {Almendros-Abad}, {K{\"o}hler}  \& {Scholz}}{{Kubiak}
  et~al.}{2021}]{Kubiak21}
{Kubiak} K.,  {Mu{\v{z}}i{\'c}} K.,  {Sousa} I.,  {Almendros-Abad} V.,
  {K{\"o}hler} R.,   {Scholz} A.,  2021, \mn@doi [\aap]
  {10.1051/0004-6361/202039899}, \href
  {https://ui.adsabs.harvard.edu/abs/2021A&A...650A..48K} {650, A48}

\bibitem[\protect\citeauthoryear{{K{\"u}pper}, {Maschberger}, {Kroupa}  \&
  {Baumgardt}}{{K{\"u}pper} et~al.}{2011}]{Kupper11}
{K{\"u}pper} A.~H.~W.,  {Maschberger} T.,  {Kroupa} P.,   {Baumgardt} H.,
  2011, \mn@doi [MNRAS] {10.1111/j.1365-2966.2011.19412.x}, \href
  {http://adsabs.harvard.edu/abs/2011MNRAS.417.2300K} {417, 2300}

\bibitem[\protect\citeauthoryear{{Lodieu}, {Hambly}, {Jameson}  \&
  {Hodgkin}}{{Lodieu} et~al.}{2008}]{Lodieu08}
{Lodieu} N.,  {Hambly} N.~C.,  {Jameson} R.~F.,   {Hodgkin} S.~T.,  2008,
  \mn@doi [MNRAS] {10.1111/j.1365-2966.2007.12676.x}, \href
  {http://adsabs.harvard.edu/abs/2008MNRAS.383.1385L} {383, 1385}

\bibitem[\protect\citeauthoryear{{Lodieu}, {Hambly}  \& {Cross}}{{Lodieu}
  et~al.}{2021}]{Lodieu21}
{Lodieu} N.,  {Hambly} N.~C.,   {Cross} N.~J.~G.,  2021, \mn@doi [\mnras]
  {10.1093/mnras/stab401}, \href
  {https://ui.adsabs.harvard.edu/abs/2021MNRAS.503.2265L} {503, 2265}

\bibitem[\protect\citeauthoryear{Low \& Lynden-Bell}{Low \&
  Lynden-Bell}{1976}]{Low76}
Low C.,  Lynden-Bell D.,  1976, MNRAS, 176, 367

\bibitem[\protect\citeauthoryear{{Lueber}, {Kitzmann}, {Bowler}, {Burgasser}
  \& {Heng}}{{Lueber} et~al.}{2022}]{Lueber22}
{Lueber} A.,  {Kitzmann} D.,  {Bowler} B.~P.,  {Burgasser} A.~J.,   {Heng} K.,
  2022, \mn@doi [\apj] {10.3847/1538-4357/ac63b9}, \href
  {https://ui.adsabs.harvard.edu/abs/2022ApJ...930..136L} {930, 136}

\bibitem[\protect\citeauthoryear{{Luhman}, {Stauffer}, {Muench}, {Rieke},
  {Lada}, {Bouvier}  \& {Lada}}{{Luhman} et~al.}{2003}]{Luhman03b}
{Luhman} K.~L.,  {Stauffer} J.~R.,  {Muench} A.~A.,  {Rieke} G.~H.,  {Lada}
  E.~A.,  {Bouvier} J.,   {Lada} C.~J.,  2003, \mn@doi [ApJ] {10.1086/376594},
  \href {http://adsabs.harvard.edu/abs/2003ApJ...593.1093L} {593, 1093}

\bibitem[\protect\citeauthoryear{{Luhman}, {Esplin}  \& {Loutrel}}{{Luhman}
  et~al.}{2016}]{Luhman16}
{Luhman} K.~L.,  {Esplin} T.~L.,   {Loutrel} N.~P.,  2016, \mn@doi [ApJ]
  {10.3847/0004-637X/827/1/52}, \href
  {http://adsabs.harvard.edu/abs/2016ApJ...827...52L} {827, 52}

\bibitem[\protect\citeauthoryear{Malmberg, Davies  \& Chambers}{Malmberg
  et~al.}{2007}]{Malmberg07a}
Malmberg D.,  Davies M.~B.,   Chambers J.~E.,  2007, MNRAS, 377, L1

\bibitem[\protect\citeauthoryear{Maschberger}{Maschberger}{2013}]{Maschberger13}
Maschberger T.,  2013, MNRAS, 429, 1725

\bibitem[\protect\citeauthoryear{Maschberger \& Clarke}{Maschberger \&
  Clarke}{2011}]{Maschberger11}
Maschberger T.,  Clarke C.~J.,  2011, MNRAS, 416, 541

\bibitem[\protect\citeauthoryear{{Mayer}, {Quinn}, {Wadsley}  \&
  {Stadel}}{{Mayer} et~al.}{2002}]{Mayer02}
{Mayer} L.,  {Quinn} T.,  {Wadsley} J.,   {Stadel} J.,  2002, \mn@doi [Science]
  {10.1126/science.1077635}, \href
  {https://ui.adsabs.harvard.edu/abs/2002Sci...298.1756M} {298, 1756}

\bibitem[\protect\citeauthoryear{{Meru} \& {Bate}}{{Meru} \&
  {Bate}}{2012}]{Meru12}
{Meru} F.,  {Bate} M.~R.,  2012, \mn@doi [\mnras]
  {10.1111/j.1365-2966.2012.22035.x}, \href
  {https://ui.adsabs.harvard.edu/abs/2012MNRAS.427.2022M} {427, 2022}

\bibitem[\protect\citeauthoryear{{Moeckel} \& {Bonnell}}{{Moeckel} \&
  {Bonnell}}{2009a}]{Moeckel09a}
{Moeckel} N.,  {Bonnell} I.~A.,  2009a, \mn@doi [MNRAS]
  {10.1111/j.1365-2966.2009.14813.x}, \href
  {http://adsabs.harvard.edu/abs/2009MNRAS.396.1864M} {396, 1864}

\bibitem[\protect\citeauthoryear{{Moeckel} \& {Bonnell}}{{Moeckel} \&
  {Bonnell}}{2009b}]{Moeckel09b}
{Moeckel} N.,  {Bonnell} I.~A.,  2009b, \mn@doi [MNRAS]
  {10.1111/j.1365-2966.2009.15499.x}, \href
  {http://adsabs.harvard.edu/abs/2009MNRAS.400..657M} {400, 657}

\bibitem[\protect\citeauthoryear{{Morii} et~al.,}{{Morii}
  et~al.}{2023}]{Morii23}
{Morii} K.,  et~al., 2023, \mn@doi [arXiv e-prints]
  {10.48550/arXiv.2304.01757}, \href
  {https://ui.adsabs.harvard.edu/abs/2023arXiv230401757M} {p. arXiv:2304.01757}

\bibitem[\protect\citeauthoryear{{Mu{\v{z}}i{\'c}}, {Scholz}, {Geers}  \&
  {Jayawardhana}}{{Mu{\v{z}}i{\'c}} et~al.}{2015}]{Muzic15}
{Mu{\v{z}}i{\'c}} K.,  {Scholz} A.,  {Geers} V.~C.,   {Jayawardhana} R.,  2015,
  \mn@doi [\apj] {10.1088/0004-637X/810/2/159}, \href
  {https://ui.adsabs.harvard.edu/abs/2015ApJ...810..159M} {810, 159}

\bibitem[\protect\citeauthoryear{Olczak, Spurzem  \& Henning}{Olczak
  et~al.}{2011}]{Olczak11}
Olczak C.,  Spurzem R.,   Henning T.,  2011, A\&A, 532, 119

\bibitem[\protect\citeauthoryear{Padoan \& Nordlund}{Padoan \&
  Nordlund}{2004}]{Padoan04}
Padoan P.,  Nordlund {\AA}.,  2004, ApJ, 617, 559

\bibitem[\protect\citeauthoryear{Parker \& {Alves de Oliveira}}{Parker \&
  {Alves de Oliveira}}{2017}]{Parker17a}
Parker R.~J.,  {Alves de Oliveira} C.,  2017, MNRAS, 468, 4340

\bibitem[\protect\citeauthoryear{Parker \& Andersen}{Parker \&
  Andersen}{2014}]{Parker14c}
Parker R.~J.,  Andersen M.,  2014, MNRAS, 441, 784

\bibitem[\protect\citeauthoryear{Parker \& Goodwin}{Parker \&
  Goodwin}{2015}]{Parker15b}
Parker R.~J.,  Goodwin S.~P.,  2015, MNRAS, 449, 3381

\bibitem[\protect\citeauthoryear{Parker \& Quanz}{Parker \&
  Quanz}{2012}]{Parker12a}
Parker R.~J.,  Quanz S.~P.,  2012, MNRAS, 419, 2448

\bibitem[\protect\citeauthoryear{Parker \& Reggiani}{Parker \&
  Reggiani}{2013}]{Parker13b}
Parker R.~J.,  Reggiani M.~M.,  2013, MNRAS, 432, 2378

\bibitem[\protect\citeauthoryear{Parker, Bouvier, Goodwin, Moraux, Allison,
  Guieu  \& G{\"u}del}{Parker et~al.}{2011}]{Parker11b}
Parker R.~J.,  Bouvier J.,  Goodwin S.~P.,  Moraux E.,  Allison R.~J.,  Guieu
  S.,   G{\"u}del M.,  2011, MNRAS, 412, 2489

\bibitem[\protect\citeauthoryear{Parker, Wright, Goodwin  \& Meyer}{Parker
  et~al.}{2014}]{Parker14b}
Parker R.~J.,  Wright N.~J.,  Goodwin S.~P.,   Meyer M.~R.,  2014, MNRAS, 438,
  620

\bibitem[\protect\citeauthoryear{Parker, Goodwin, Wright, Meyer  \&
  Quanz}{Parker et~al.}{2016}]{Parker16c}
Parker R.~J.,  Goodwin S.~P.,  Wright N.~J.,  Meyer M.~R.,   Quanz S.~P.,
  2016, MNRAS, 459, L119

\bibitem[\protect\citeauthoryear{{Pearson}, {Scholz}, {Teixeira},
  {Mu{\v{z}}i{\'c}}  \& {Eisl{\"o}ffel}}{{Pearson} et~al.}{2020}]{Pearson20}
{Pearson} S.,  {Scholz} A.,  {Teixeira} P.~S.,  {Mu{\v{z}}i{\'c}} K.,
  {Eisl{\"o}ffel} J.,  2020, \mn@doi [\mnras] {10.1093/mnras/staa2997}, \href
  {https://ui.adsabs.harvard.edu/abs/2020MNRAS.499.2292P} {499, 2292}

\bibitem[\protect\citeauthoryear{{Plunkett}, {Fern{\'a}ndez-L{\'o}pez}, {Arce},
  {Busquet}, {Mardones}  \& {Dunham}}{{Plunkett} et~al.}{2018}]{Plunkett18}
{Plunkett} A.~L.,  {Fern{\'a}ndez-L{\'o}pez} M.,  {Arce} H.~G.,  {Busquet} G.,
  {Mardones} D.,   {Dunham} M.~M.,  2018, \mn@doi [\aap]
  {10.1051/0004-6361/201732372}, \href
  {http://adsabs.harvard.edu/abs/2018A%26A...615A...9P} {615, A9}

\bibitem[\protect\citeauthoryear{{Portegies Zwart}, Makino, McMillan  \&
  Hut}{{Portegies Zwart} et~al.}{1999}]{Zwart99}
{Portegies Zwart} S.~F.,  Makino J.,  McMillan S. L.~W.,   Hut P.,  1999, A\&A,
  348, 117

\bibitem[\protect\citeauthoryear{{Portegies Zwart}, McMillan, Hut  \&
  Makino}{{Portegies Zwart} et~al.}{2001}]{Zwart01}
{Portegies Zwart} S.~F.,  McMillan S. L.~W.,  Hut P.,   Makino J.,  2001,
  MNRAS, 321, 199

\bibitem[\protect\citeauthoryear{Prim}{Prim}{1957}]{Prim57}
Prim R.~C.,  1957, Bell Syst. Tech. J., 36, 1389

\bibitem[\protect\citeauthoryear{Rees}{Rees}{1976}]{Rees76}
Rees M.~J.,  1976, MNRAS, 176, 483

\bibitem[\protect\citeauthoryear{Salpeter}{Salpeter}{1955}]{Salpeter55}
Salpeter E.~E.,  1955, ApJ, 121, 161

\bibitem[\protect\citeauthoryear{S{\'a}nchez \& Alfaro}{S{\'a}nchez \&
  Alfaro}{2009}]{Sanchez09}
S{\'a}nchez N.,  Alfaro E.~J.,  2009, ApJ, 696, 2086

\bibitem[\protect\citeauthoryear{{Schmeja} \& {Klessen}}{{Schmeja} \&
  {Klessen}}{2006}]{Schmeja06}
{Schmeja} S.,  {Klessen} R.~S.,  2006, \mn@doi [A\&A]
  {10.1051/0004-6361:20054464}, 449, 151

\bibitem[\protect\citeauthoryear{{Schmidt-Kaler}}{{Schmidt-Kaler}}{1982}]{Schmidt82}
{Schmidt-Kaler} T.,  1982, Bulletin d'Information du Centre de Donnees
  Stellaires, \href {http://adsabs.harvard.edu/abs/1982BICDS..23....2S} {23, 2}

\bibitem[\protect\citeauthoryear{{Schoettler}, {de Bruijne}, {Vaher}  \&
  {Parker}}{{Schoettler} et~al.}{2020}]{Schoettler20}
{Schoettler} C.,  {de Bruijne} J.,  {Vaher} E.,   {Parker} R.~J.,  2020,
  \mn@doi [\mnras] {10.1093/mnras/staa1228}, \href
  {https://ui.adsabs.harvard.edu/abs/2020MNRAS.tmp.1474S} {495, 3104}

\bibitem[\protect\citeauthoryear{{Scholz}, {Muzic}, {Jayawardhana}, {Quinlan}
  \& {Wurster}}{{Scholz} et~al.}{2022}]{Scholz22}
{Scholz} A.,  {Muzic} K.,  {Jayawardhana} R.,  {Quinlan} L.,   {Wurster} J.,
  2022, \mn@doi [\pasp] {10.1088/1538-3873/ac9431}, \href
  {https://ui.adsabs.harvard.edu/abs/2022PASP..134j4401S} {134, 104401}

\bibitem[\protect\citeauthoryear{{Scholz}, {Muzic}, {Jayawardhana},
  {Almendros-Abad}  \& {Wilson}}{{Scholz} et~al.}{2023}]{Scholz23}
{Scholz} A.,  {Muzic} K.,  {Jayawardhana} R.,  {Almendros-Abad} V.,   {Wilson}
  I.,  2023, \mn@doi [\aj] {10.3847/1538-3881/acc65d}, \href
  {https://ui.adsabs.harvard.edu/abs/2023AJ....165..196S} {165, 196}

\bibitem[\protect\citeauthoryear{{Segura-Cox} et~al.,}{{Segura-Cox}
  et~al.}{2020}]{SeguraCox20}
{Segura-Cox} D.~M.,  et~al., 2020, \mn@doi [\nat] {10.1038/s41586-020-2779-6},
  \href {https://ui.adsabs.harvard.edu/abs/2020Natur.586..228S} {586, 228}

\bibitem[\protect\citeauthoryear{{Smullen}, {Kratter}  \& {Shannon}}{{Smullen}
  et~al.}{2016}]{Smullen16}
{Smullen} R.~A.,  {Kratter} K.~M.,   {Shannon} A.,  2016, \mn@doi [\mnras]
  {10.1093/mnras/stw1347}, \href
  {https://ui.adsabs.harvard.edu/abs/2016MNRAS.461.1288S} {461, 1288}

\bibitem[\protect\citeauthoryear{{Spera}, {Mapelli}  \& {Jeffries}}{{Spera}
  et~al.}{2016}]{Spera16}
{Spera} M.,  {Mapelli} M.,   {Jeffries} R.~D.,  2016, \mn@doi [\mnras]
  {10.1093/mnras/stw998}, \href
  {https://ui.adsabs.harvard.edu/abs/2016MNRAS.460..317S} {460, 317}

\bibitem[\protect\citeauthoryear{{Spitzer}}{{Spitzer}}{1969}]{Spitzer69}
{Spitzer} Jr. L.,  1969, \mn@doi [ApJL] {10.1086/180451}, \href
  {http://adsabs.harvard.edu/abs/1969ApJ...158L.139S} {158, L139}

\bibitem[\protect\citeauthoryear{Stamatellos \& Whitworth}{Stamatellos \&
  Whitworth}{2009}]{Stamatellos09}
Stamatellos D.,  Whitworth A.~P.,  2009, MNRAS, 392, 413

\bibitem[\protect\citeauthoryear{Thies \& Kroupa}{Thies \&
  Kroupa}{2008}]{Thies08}
Thies I.,  Kroupa P.,  2008, MNRAS, 390, 1200

\bibitem[\protect\citeauthoryear{{Trenti} \& {van der Marel}}{{Trenti} \& {van
  der Marel}}{2013}]{Trenti13}
{Trenti} M.,  {van der Marel} R.,  2013, \mn@doi [MNRAS]
  {10.1093/mnras/stt1521}, \href
  {http://adsabs.harvard.edu/abs/2013MNRAS.435.3272T} {435, 3272}

\bibitem[\protect\citeauthoryear{{Veras} \& {Raymond}}{{Veras} \&
  {Raymond}}{2012}]{Veras12}
{Veras} D.,  {Raymond} S.~N.,  2012, \mn@doi [\mnras]
  {10.1111/j.1745-3933.2012.01218.x}, \href
  {https://ui.adsabs.harvard.edu/abs/2012MNRAS.421L.117V} {421, L117}

\bibitem[\protect\citeauthoryear{{Ward-Duong} et~al.,}{{Ward-Duong}
  et~al.}{2015}]{Ward-Duong15}
{Ward-Duong} K.,  et~al., 2015, MNRAS, \href
  {http://adsabs.harvard.edu/abs/2015MNRAS.449.2618W} {449, 2618}

\bibitem[\protect\citeauthoryear{{Whitworth} \& {Stamatellos}}{{Whitworth} \&
  {Stamatellos}}{2006}]{Whitworth06}
{Whitworth} A.~P.,  {Stamatellos} D.,  2006, \mn@doi [\aap]
  {10.1051/0004-6361:20065806}, \href
  {https://ui.adsabs.harvard.edu/abs/2006A&A...458..817W} {458, 817}

\bibitem[\protect\citeauthoryear{{Zapatero Osorio} et~al.,}{{Zapatero Osorio}
  et~al.}{2014}]{Osorio14b}
{Zapatero Osorio} M.~R.,  et~al., 2014, \mn@doi [A\&A]
  {10.1051/0004-6361/201424634}, \href
  {https://ui.adsabs.harvard.edu/abs/2014A&A...572A..67Z} {572, A67}

\bibitem[\protect\citeauthoryear{{Zheng}, {Kouwenhoven}  \& {Wang}}{{Zheng}
  et~al.}{2015}]{Zheng15}
{Zheng} X.,  {Kouwenhoven} M.~B.~N.,   {Wang} L.,  2015, \mn@doi [MNRAS]
  {10.1093/mnras/stv1832}, \href
  {https://ui.adsabs.harvard.edu/abs/2015MNRAS.453.2759Z} {453, 2759}

\makeatother
\end{thebibliography}

\label{lastpage}

\end{document}